\documentclass[journal]{IEEEtran}
\usepackage[boxruled]{algorithm2e}
\usepackage{cite}
\usepackage{graphicx}
\usepackage{psfrag}
\usepackage{subfigure}
\usepackage{url}
\usepackage{amsmath}
\usepackage{array}
\usepackage{amssymb}
\usepackage{amsfonts}
\usepackage{graphicx}
\usepackage{epstopdf}
\usepackage{algorithmic}

\newtheorem{lemma}{Lemma}

\newtheorem{corollary}{Corollary}
\newtheorem{remark}{Remark}
\newtheorem{definition}{Definition}
\newtheorem{theorem}{Theorem}

\newtheorem{example}{Example}

\usepackage{setspace}
\usepackage{amscd}
\usepackage{mathrsfs}
\usepackage{epsfig}
\usepackage{color}
\usepackage{textcomp}
\usepackage{multirow}
\usepackage{threeparttable}

\title{Wireless Bidirectional Relaying, Latin Squares and Graph Vertex Coloring}
\begin{document}

\author{Vijayvaradharaj T. Muralidharan and B. Sundar Rajan$^{*}$,~\IEEEmembership{Senior Member,~IEEE}\thanks{The authors are with the Dept. of Electrical Communication Engineering, Indian Institute of Science, Bangalore-560012, India (e-mail:tmvijay@ece.iisc.ernet.in; bsrajan@ece.iisc.ernet.in).}}

\maketitle
\begin{abstract}
The problem of obtaining network coding maps for the physical layer network coded two-way relay channel is considered, using the denoise-and-forward forward protocol. It is known that network coding maps used at the relay node which ensure unique decodability at the end nodes form Latin Squares. Also, it is known that minimum distance of the effective constellation at the relay node becomes zero, when the ratio of the fade coefficients from the end node to the relay node, belongs to a finite set of complex numbers called the singular fade states, determined by the signal set used. Furthermore, it has been shown recently that the problem of obtaining network coding maps which remove the harmful effects of singular fade states, reduces to the one of obtaining Latin Squares, which satisfy certain constraints called \textit{singularity removal constraints}. In this paper, it is shown that the singularity removal constraints along with the row and column exclusion conditions of a Latin Square, can be compactly represented by a graph called the \textit{singularity removal graph} determined by the singular fade state and the signal set used. It is shown that a Latin Square which removes a singular fade state can be obtained from a proper vertex coloring of the corresponding singularity removal graph. The earlier known approach of completing partially filled Latin Squares to obtain the network coding maps, can be viewed as a special way of vertex coloring of the singularity removal graph. The minimum number of symbols used to fill in a Latin Square which removes a singular fade state is equal to the chromatic number of the singularity removal graph. It is shown that for any square $M$-QAM signal set, there exists singularity removal graphs whose chromatic numbers exceed $M$ and hence require more than $M$ colors for vertex coloring. Also, it is shown that for any $2^{\lambda}$-PSK signal set, $\lambda \geq 3,$ all the singularity removal graphs can be colored using $M=2^{\lambda}$ colors.
\end{abstract}
\section{Introduction and Background}
Network coding has emerged as an attractive alternative to routing because of the throughput improvement it provides by reducing the number of channel uses. 
In a wireless scenario, further throughput improvement can be achieved through physical layer network coding, a technique in which nodes are allowed to transmit simultaneously, instead of transmitting in orthogonal slots. 

Information theoretic and communication theoretic benefits of physical layer network coding have been studied widely in the literature \cite{ZLL,NaGa,HeN,SyBu,APT1}.
The concept of physical layer network coding was first introduced in \cite{ZLL}, in which it was observed that allowing the two users in a two-way relay channel to transmit simultaneously results in throughput improvement, without a degradation in the error performance. The compute and forward framework was introduced in \cite{NaGa}, in which it was shown that allowing the intermediate relay nodes to decode a linear combination of the messages according to the channel coefficients results in significantly higher achievable rates. A multi-level coding scheme based on the compute and forward framework was proposed in \cite{HeN}. The constellation constrained achievable rate regions for the wireless two-way relay channel with a layered hierarchical code design were obtained in \cite{SyBu}. In \cite{APT1}, it was shown that for uncoded communication, changing the network coding map adaptively according to the channel condition results in significant increase in throughput. In \cite{VNR}, such adaptive network coding maps were obtained using Latin Squares. Symbol error rate performance analysis of the physical layer network coding scheme in which network coding maps are changed adaptively according to the channel condition was presented in \cite{VvR_Perf}. Obtaining Latin Square based network coding maps for the MIMO two-way relaying scenario was considered in \cite{VvR_MIMO}. Physical layer network coding using Latin Squares for two-way relaying with QAM signal set was considered in \cite{NVR}.  

\begin{figure}[htbp]
\centering
\subfigure[MA Phase]{
\includegraphics[totalheight=.4in,width=1.25in]{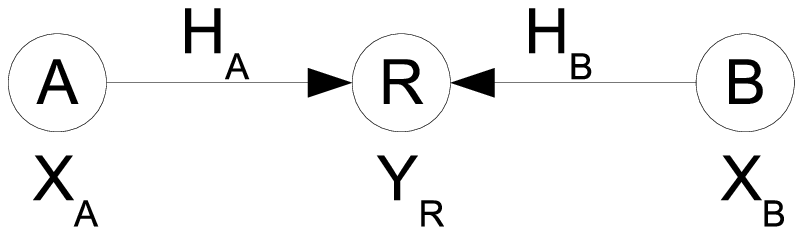}
\label{fig:phase1}
}

\subfigure[BC Phase]{
\includegraphics[totalheight=.4in,width=1.25in]{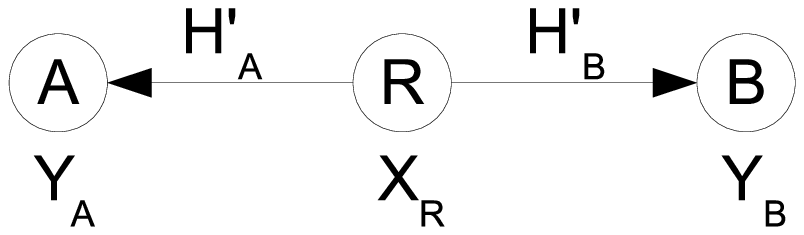}
\label{fig:phase2}
}
\caption{The Two Way Relay Channel}
\label{two_way_relay}
\end{figure}

In this paper, we consider the design of physical layer network coding maps for the wireless two-way relay channel shown in Fig. \ref{two_way_relay}, which employs the denoise-and-forward protocol. The end nodes A and B exchange messages with the help of the relay node R. We consider the framework introduced in \cite{APT1}, in which the network coding map used at R is chosen based on the channel realization. All the nodes are assumed to be be half-duplex. Communication takes place in two phases: (i) Multiple Access (MA) phase during which A and B transmit simultaneously to R and (ii) Broadcast (BC) phase during which R transmits the estimate of a many-to-one function of A's and B's transmission. To ensure decodability of B's (A's) message at node A (B), the many-to-one function used at R should satisfy a condition called exclusive law \cite{APT1} and all maps satisfying exclusive law form Latin Squares\cite{VNR}.  

Due to the simultaneous transmission of nodes A and B during the MA phase, distance shortening occurs in the effective constellation at R, when the ratio of the channel fade coefficients of the B-R and A-R links called the fade state, falls in the neighbourhood of a set of complex numbers called the singular fade states\cite{APT1,VNR}. The problem of finding network coding maps which remove the harmful effects of these singular fade states, henceforth referred as the removal of singular fade states, reduces to finding Latin Squares which satisfy certain constraints called the singularity removal constraints\cite{VNR}. In \cite{VNR}, such Latin Squares were obtained by completing partially filled Latin Squares, obtained after combining singularity removal constraints (more details on this are provided in Section III-B). Also, it was conjectured in \cite{VNR}, that for any $2^{\lambda}$-PSK signal set, the singular fade states can be removed by Latin Squares which are filled in with the minimum number of symbols $2^{\lambda}.$ Note that the number of symbols used in the Latin Square is equal to the size of the signal set used during the BC phase and hence needs to be kept as much minimum as possible \cite{VNR}. 

This paper focusses on obtaining Latin Squares which remove singular fade states using graph vertex coloring. The approach in \cite{VNR} which involves the completion of partially filled Latin Squares can be viewed as a special case.  

The preliminary notations related to graph theory and the signal model are presented in Section II-A and Section II-B respectively. The contributions and organization of the rest of the paper are as follows:
\begin{itemize}
\item
The notion of singularity removal graph is introduced, which is a compact representation of the singularity removal constraints as well as the constraints imposed by the exclusive law. It is shown that every Latin Square which removes a singular fade state can be obtained by a proper vertex coloring (for definition, see Section II A) of the corresponding singularity removal graph (Section III-A).  
\item
The two stage approach adopted in \cite{VNR} which involves completion of partially filled Latin Squares obtained by combining singularity removal constraints can be viewed as special way of coloring the singularity removal graph. The two stage approach can be viewed equivalently as first coloring a subgraph of the singularity removal graph, referred to as the vital subgraph, before coloring the rest of the singularity removal graph. The drawback of this two stage approach in relation to the one stage approach of coloring the entire singularity removal graph is that the two stage approach may not always result in a Latin Square with the minimum number of symbols (Section III-B).
\item
The minimum number of symbols in a Latin Square which removes a singular fade state is equal to the chromatic number (for definition, see Section II A) of the corresponding singularity removal graph. If for a singular fade state, the chromatic number of the singularity removal graph exceeds the size of the signal set $M$ used at the end nodes, then it means that there does not exist a Latin Square with the minimum number of symbols $M$ which removes that singular fade state. It is shown that for any  square QAM signal set, there exists singular fade states for which the singularity removal graphs have chromatic numbers more than $M$ (Section IV).
\item
The conjecture in \cite{VNR} that for any $2^{\lambda}$-PSK signal set, all the singular fade states can be removed using Latin Squares which contain only $2^{\lambda}$ symbols is proved. In other words, it is shown that for any $2^{\lambda}$-PSK signal set, all the singularity removal graphs can be colored using only $2^{\lambda}$ colors (Section V).
\end{itemize}

\section{Preliminaries}
\subsection{Graph Vertex Coloring}
In this subsection, some notations and definitions related to graph vertex coloring are presented. For a detailed treatment of graph coloring, see \cite{We,Di}.

Let $G$ be a simple graph (an undirected graph without loops and without multiple edges between vertices) with vertex set $V(G)$ and edge set $E(G).$ The set $E(G)$ contains unordered pairs from $V(G) \times V(G)$ and two vertices $v_1$ and $v_2$ are said to be adjacent if $(v_1,v_2)\in E(G).$ A clique is a subgraph of $G$ in which every pair of vertices are adjacent. Clique number of $G,$ denoted by $\omega(G)$ is the number of vertices in a clique of $G$ which contains the maximum number of vertices.

A $k$-coloring of a graph is an onto map $f: V(G) \rightarrow S,$ where $S$ is the set of colors with $\vert S \vert=k$ and $f(v)$ is said to be the color assigned to the vertex $v.$ A $k$-coloring is said to be proper if no two adjacent vertices are assigned the same color. A graph is said to be $k$-colorable if it has a proper $k$-coloring. The minimum value of $k$ for which a proper $k$-coloring exists for $G$ is called the chromatic number of $G$ denoted by $\chi(G)$ and $G$ is said to be $k$-chromatic. Throughout the paper we consider only proper vertex colorings of graphs and by $k$-coloring what we actually mean is a proper $k$-coloring. A $k$-coloring for a graph $G$ is said to be optimal if $\chi(G)=k.$
\subsection{Wireless two-way relaying}
In this subsection, the scheme proposed in \cite{VNR} for wireless two-way relaying is briefly described along with some useful definitions.

Throughout, the denoise-and-forward protocol for two-way relaying is considered which involves two phases: MA phase and BC phase. During the MA phase, end nodes A and B which want to exchange information, simultaneously transmit signal points from a signal set $\mathcal{S}$ of cardinality $M.$ The received signal at R during MA phase is given by,
$y_R=H_A x_A+H_B x_B+n_R,$
where $H_A$ and $H_B$ are the fading coefficients associated with the A-R and B-R links, $x_A,x_B \in \mathcal{S}$ are the complex symbols transmitted by A and B respectively, and $n_R \sim \mathcal{CN}(0,\sigma^2)$ is the additive noise at R, where $\mathcal{CN}(0,\sigma^2)$ denotes a circularly symmetric complex Gaussian random variable with mean zero and variance equal to $\sigma^2.$ The ratio $H_B/H_A$ is called the fade state and is denoted by $\gamma e^{j \theta}.$

During the BC phase, R transmits a many-to-one function of the joint ML estimate $(\hat{x}_A,\hat{x}_B)$ of $(x_A,x_B).$ Let $f^{(\gamma,\theta)}:\mathcal{S}\times \mathcal{S}\rightarrow\mathcal{S}' $ denote the many-to-one network coding map used at R, which is chosen based on the fade state $\gamma e^{j \theta},$ where $\mathcal{S}'$ whose cardinality lies between $M$ and $M^2$ is the signal set used during the BC phase. The received signal at node $J\in \{A,B\}$ during the BC phase is given by, $y_J=H'_J f^{(\gamma,\theta)}(\hat{x}_A,\hat{x}_B)+n_J,$ where $H'_J$ is the fading coefficient of the $R-J$ link, and $n_J \sim \mathcal{CN}(0,\sigma^2)$ is the additive noise at node $J.$ To ensure that A (B) is able to decode B's (A's) message, given the knowledge of its own message, the network coding map needs to satisfy the exclusive law \cite{APT1},
\begin{align*}
f^{(\gamma,\theta)}(x_A,x_B) \neq f^{(\gamma,\theta)}(x'_A,x_B), \forall x_A \neq x'_A, x_A, x'_A,x_B \in \mathcal{S},\\
f^{(\gamma,\theta)}(x_A,x_B) \neq f^{(\gamma,\theta)}(x_A,x'_B), \forall x_B \neq x'_B, x_A, x_B,x'_B \in \mathcal{S}.
\end{align*}

The conditions imposed by exclusive law are nothing but the row and column exclusion conditions of a Latin Square, i.e., the maps satisfying exclusive law form Latin Squares \cite{VNR}. Note that a Latin Square is an $M \times M$ array with $t \geq M$ symbols, with exactly one symbol filled in a cell of the array, in such a way that no two symbols repeat in any row as well as any column. The points in $\mathcal{S}$ are assigned labels from 1 to $M.$ The row index of the Latin Square denotes A's transmission, the column index denotes B's transmission and the symbols filled in the Latin Square denote R's transmission. Note that $\vert \mathcal{S}' \vert$ should be at least $M$ to ensure the satisfaction of exclusive law.

The set $\{H_A x_A +H_B x_B : x_A,x_B \in \mathcal{S}\}$ is the called the effective constellation at R. The values of $\gamma e^{j \theta}$ for which the minimum distance of the effective constellation at R becomes zero are called singular fade states \cite{VNR}. Alternatively, a value of fade state is a singular fade state if the effective constellation at R has less than $M^2$ points.

\begin{example}
\label{example1}
When nodes A and B use 4-QAM signal set $\{\pm 1\pm j\},$ for $\gamma e^{j \theta}=\frac{(1+j)}{2},$ the effective constellation at R normalized by $H_A,$ has 12 points given by $\{\pm 1,\pm j,\pm 2 \pm j, \pm 1 \pm 2j\}.$ Hence, $\gamma e^{j \theta}=\frac{(1+j)}{2}$ is a singular fade state for 4-QAM signal set.
\end{example}

When $\gamma e^{j \theta}$ is a singular fade state, to eliminate the harmful effect of the fade state falling in the neighbourhood of $\gamma e^{j \theta},$ the map $f^{(\gamma,\theta)}$ should satisfy the condition that $f^{(\gamma,\theta)}(x_A,x_B)=f^{(\gamma,\theta)}(x'_A,x'_B),$ for all $(x_A,x_B)$ and $(x'_A,x'_B)$ for which $x_A+\gamma e^{j \theta}x_B=x'_A+\gamma e^{j \theta}x'_B.$ A map which satisfies this condition is said to remove the singular fade state $\gamma e^{j \theta}.$ The removal of a singular fade state requires certain cells in the Latin Square which removes it to be filled in with the same symbol.
\begin{definition}
A singularity removal constraint for a singular fade state $s$ and signal set $\mathcal{S}$ is a subset of $\mathcal{S}\times \mathcal{S},$ denoted by $c(s)$ which satisfies the following conditions: 
\begin{itemize}
\item
$ x_A+sx_B=x'_A+sx'_B,\forall (x_A,x_B), (x'_A,x'_B) \in c(s),$
\item
$  x_A+sx_B \neq x'_A+sx'_B,\forall (x_A,x_B)\in c(s), (x'_A,x'_B)\notin c(s).$
\end{itemize}

Let $\mathcal{C}(s)$ denote the set of all singularity removal constraints, for the singular fade state $s.$ 
\end{definition}  

In a Latin Square which removes the singular fade state $s,$ all the cells which belong to a singularity removal constraint need to be filled in with the same symbol.
For simplicity, let $c_i, i \in \{1,2,\dotso, \vert \mathcal{C}(s) \vert \},$ denote the $i$th singularity removal constraint. 

\begin{example}
\label{example2}
Continuing with Example \ref{example1}, consider the singular fade state $(1+j)/2$ for 4-QAM signal set. Let the points in 4-QAM signal set be labelled from 1 to 4 in the order $\{ -1-j,  -1 + j,   1 - j ,  1 + j\}.$ Since $(-1-j)+(1+j)/2\times(1-j)= (1-j)+(1+j)/2\times(-1+j)= - j$ and for no other $(x_A,x_B)$, the value of $x_A+\frac{(1+j)}{2} x_B$ equals $-j,$ the set $\{(1,3),(3,2)\}$ is a singularity removal constraint. It can be verified that the set of singularity removal constraints
 $\mathcal{C}\left(\frac{1+j}{2}\right)$ is given by, 

{\footnotesize
\begin{align*} 
 \{&\underbrace{\{(1,3),(3,2)\}}_{c_1},\underbrace{\{(1,4),(2,1)\}}_{c_2},\underbrace{\{(2,3),(4,2)\}}_{c_3},\underbrace{\{(3,4),(4,1)\}}_{c_4}\\
 &\hspace{-.1 cm}\underbrace{\{(1,1)\}}_{c_5},\underbrace{\{(1,2)\}}_{c_6},\underbrace{\{(2,2)\}}_{c_7},\underbrace{\{(2,4)\}}_{c_8}, \underbrace{\{(3,1)\}}_{c_9},\underbrace{\{(3,3)\}}_{c_{10}},\underbrace{\{(4,3)\}}_{c_{11}},\underbrace{\{(4,4)\}}_{c_{12}}\}.
 \end{align*}
 }
\end{example}

\begin{definition}
A constrained partial Latin Square for a singular fade state $s,$ denoted by $\mathcal{L}(s)$ is a partially filled square in which the cells which belong to a singularity removal constraint $c_i$ with $\vert c_i \vert \geq 2$ are filled in with symbol $i$ and all other cells are empty. 
\end{definition}

\begin{example}
Continuing with Example \ref{example2}, the constrained partial Latin Square $\mathcal{L}\left(\frac{1+j}{2}\right)$ is as given in Fig. \ref{CPLS_example}.
\begin{figure}
\centering
\begin{tabular}{|c|c|c|c|}
\hline  &  & 1 &2 \\ 
\hline 2 &  &3  & \\ 
\hline &1  &  &4 \\ 
\hline 4 & 3 &  & \\ 
\hline 
\end{tabular}
\caption{The constrained partial Latin Square $\mathcal{L}\left(\frac{1+j}{2}\right).$}
\label{CPLS_example}
\end{figure}
\end{example}

\begin{definition}
A Latin Square is said to remove a singular fade state if all its cells which belong to the same singularity removal constraint are filled in with the same symbol.
\end{definition}

\begin{example}
Continuing with Example \ref{example2}, the Latin Square given in Fig. \ref{Latin_srg} removes the singular fade state $(1+j)/2$ for 4-QAM signal set.
\begin{figure}
\centering
\begin{tabular}{|c|c|c|c|}
\hline  3&  5& 1 &2 \\ 
\hline 2 &4  &3  &5 \\ 
\hline 5&1  &2  &4 \\ 
\hline 4 & 3 & 5 &1 \\ 
\hline 
\end{tabular}
\caption{Latin Square which removes the singular fade state ${\frac{(1+j)}{2}}.$}
\label{Latin_srg}
\end{figure}
\end{example}

Since the number of symbols in the Latin Square equals the cardinality of the signal set used during the BC phase, among the Latin Squares which remove a singular fade state, one needs to choose a Latin Square which contains the minimum number of symbols. A particular realization of $\gamma e^{j \theta}$ need not be a singular fade state. Two criteria have been proposed in \cite{VNR} to choose one among the Latin Squares which remove singular fade states depending on $\gamma e^{j \theta}.$ Both the criteria avoid the harmful effect of distance shortening in the neighbourhood of singular fade states. The choice of the Latin Square chosen for a given realization of $\gamma e^{j\theta}$ is indicated by R to A and B using overhead bits. For more details, see \cite{VNR}.

\section{Obtaining Latin Squares which Remove Singular Fade States: A Graph Coloring Approach}
In Section III-A, the notion of singularity removal graph for a singular fade state is introduced. It is shown that a Latin Square which removes a singular fade state can be obtained by finding a proper vertex coloring of the singularity removal graph. In Section III-B, a discussion is presented related to the connection between the graph coloring approach and the procedure followed in \cite{VNR} to obtain Latin Squares which remove singular fade states. 
\subsection{Singularity Removal Graph}
The singularity removal graph for a singular fade state is defined as follows.
\begin{definition}
\label{defn_srg}
The singularity removal graph  for a singular fade state $s,$ denoted by $\mathcal{G}_{s}$ is a simple graph whose vertex set $V(\mathcal{G}_s)$ is $\{1,2,\dotso, \vert\mathcal{C}(s)\vert\}.$ An edge exists between two vertices $i$ and $j$ in $\mathcal{G}_{s}$ if and only if there exists $(x_A,x_B)\in c_i$ and $(x'_A,x'_B) \in c_j$ for which either $x_A=x'_A$ or $x_B=x'_B.$   
\end{definition}
\begin{remark}
The singularity removal graph  for a singular fade state $s$ also depends on the signal set $\mathcal{S}.$ For simplicity, $\mathcal{S}$ is not included in the notation $\mathcal{G}_{s}$ used for a singularity removal graph. 
\end{remark}
\begin{example}
\label{example4}
Continuing with Example \ref{example2}, for $s=\frac{1+j}{2},$ the singularity removal graph $\mathcal{G}_{\frac{(1+j)}{2}}$ is as shown in Fig. \ref{fig:srg_example}. In $\mathcal{G}_{\frac{(1+j)}{2}},$ an edge exists between vertex 1 and vertex 11 since $(1,3) \in c_{1}$ and $(4,3)\in c_{11}.$ Similarly, it can be verified that for all other edges, the condition given in Definition \ref{defn_srg} under which an edge exists between two vertices is satisfied.
\begin{figure}[htbp]
\centering
\includegraphics[totalheight=2.5in,width=3.5in]{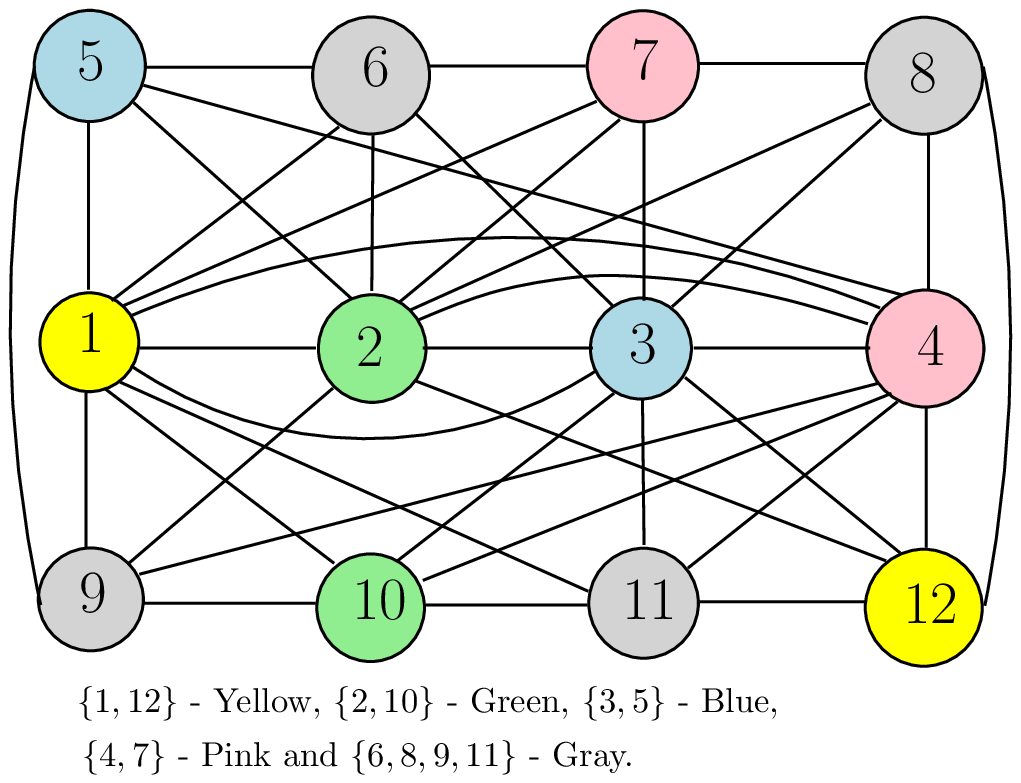}
\caption{The singularity removal graph $\mathcal{G}_{\frac{(1+j)}{2}}.$}
\label{fig:srg_example}
\end{figure}
\end{example}

Recall that in a Latin Square which removes a singular fade state $s,$ the cells which belong to the same singularity removal constraint are filled in with the same symbol . From a proper $k$ coloring of the singularity removal graph $\mathcal{G}_{s},$ a Latin Square which removes the singular fade state $s$ can be obtained as follows: Assign distinct labels from the set $\{1,2,\dotso,k\}$ to the $k$ colors used. Start with an empty Latin Square and fill in the cells which belong to the constraint $c_i$ with the color assigned to the node $i$ in $\mathcal{G}_{s}$ to obtain a Latin Square which removes the singular fade state $s.$

\begin{example}
\label{example6}
Continuing with Example \ref{example4}, a proper 5-coloring for $\mathcal{G}_{\frac{(1+j)}{2}}$ is shown in Fig. \ref{fig:srg_example}. Let 1, 2, 3, 4 and 5 be the labels assigned to the colors yellow, green, blue, pink and gray respectively. Starting with an empty Latin Square, since the nodes 1 and 12 are colored with yellow, the cells which belong to $c_1$ ($(1,3),(3,2)$) and $c_{12}$ ($(4,4)$) are filled in with the symbol 1. In a similar way, the rest of the Latin Square can be filled to obtained the Latin Square given in Fig. \ref{Latin_srg}, which removes the singular fade state $\frac{(1+j)}{2}$ for 4-QAM signal set. 
\end{example} 

Note that the approach of obtaining Latin Squares which remove singular fade states using graph vertex coloring is valid for any arbitrary signal set. Specific examples of vertex colorings for well known signal sets like QAM and PSK are considered in Section IV and Section V respectively.
For any arbitrary signal set, a Latin Square with the minimum number of symbols which removes a singular fade state can be obtained by finding an optimal vertex coloring for the corresponding singularity removal graph. Several methods have been proposed in the literature to obtain an optimal vertex coloring for a graph \cite{CoGr,Br,MaMoTo}. The problem of finding an optimal vertex coloring for a general graph is NP-complete and polynomial time algorithms which solve this problem for a general graph are not known. On the other hand, several algorithms based on sequential coloring \cite{Kl}, tabu search \cite{WaKl}, iterated local search \cite{CaDe} etc. have been proposed in the literature which can provide near-optimal solutions. Also, for certain classes of graphs, linear and polynomial time algorithms for graph vertex coloring are known \cite{web_1}. 
With regards to the problem of wireless bidirectional relaying, it is not still known whether the class of singularity removal graphs can be colored using a polynomial time algorithm for any arbitrary signal set. For a comprehensive bibliography on graph vertex coloring, see \cite{web_2}. 
  
A Latin Square which removes the singular fade state $s$ which is filled in with $k$ symbols exists if and only if there exists a $k$-coloring for $\mathcal{G}_{s}.$
In other words, the minimum number of symbols in a Latin Square which removes a singular fade state $s$ is equal to the chromatic number $\chi(\mathcal{G}_{s})$ of the singularity removal graph $\mathcal{G}_{s}.$ Note that the satisfaction of exclusive law necessitates the use of at least $M$ symbols in an $M \times M$ Latin Square. A Latin Square which removes a singular fade state $s$ and which contains only $M$ symbols exists if and only if $\chi(\mathcal{G}_{s})=M.$ More details related to this are presented in Section IV.    

\subsection{Relationship between the graph coloring approach and the approach used in \cite{VNR}}
In this subsection we discuss the connection between the graph coloring approach to obtain Latin Squares which remove singular fade states, introduced in the previous subsection and the two step procedure followed in \cite{VNR}. A brief description of the procedure followed in \cite{VNR} to obtain Latin Squares which remove singular fade states is first provided.

The procedure followed in \cite{VNR} involves two steps. Note that the constrained partial Latin Square can contain more than $M$ symbols. Since it is desirable to obtain a Latin Square which is filled in with the minimum number of symbols, in the first step, some symbols in the constrained partial Latin Square are combined to obtain a partially filled Latin Square which contains less than or equal to $M$ symbols, if possible. This step is illustrated in the following example.
\begin{example}
\label{example7}
Consider 4-PAM signal set $\{-3,-1,1,3\}$ labelled by the integers $\{1,2,3,4\}.$ It can be verified that $-2$ is a singular fade state. For this case, the set of singularity removal constraints is given by, 

{\footnotesize
\begin{align*} 
 \{&\underbrace{\{(1,1),(3,2)\}}_{c_1},\underbrace{\{(1,2),(3,3)\}}_{c_2},\underbrace{\{(1,3),(3,4)\}}_{c_3},\underbrace{\{(2,1),(3,2)\}}_{c_4}\\
 &\hspace{-.1 cm}\underbrace{\{(2,2),(4,3)\}}_{c_5},\underbrace{\{(2,3),(4,4)\}}_{c_6},\underbrace{\{(1,4)\}}_{c_7},\underbrace{\{(2,4)\}}_{c_8}, \underbrace{\{(3,1)\}}_{c_9},\underbrace{\{(4,1)\}}_{c_{10}}\}.
 \end{align*}
 }The constrained partial Latin Square $\mathcal{L}(-2)$ is given in Fig. \ref{CPLS_2}. Let the symbols 1 and 6 be combined, with 1 denoting the new symbol. Also, let 3 and 4 be combined, with 3 denoting the new symbol and replace the symbol 5 in $\mathcal{L}(-2)$ with the symbol 4. After this, we get a partially filled Latin Square shown in Fig. \ref{PFLS_2}.
\begin{figure}
\centering
\begin{tabular}{|c|c|c|c|}
\hline  1& 2 & 3 & \\ 
\hline 4 &  5&6  & \\ 
\hline &1  &2  &3 \\ 
\hline  & 4 & 5 &6 \\ 
\hline 
\end{tabular}
\caption{The constrained partial Latin Square $\mathcal{L}\left(-2\right).$}
\label{CPLS_2}
\end{figure}
\begin{figure}
\centering
\begin{tabular}{|c|c|c|c|}
\hline  1& 2 & 3 & \\ 
\hline 3 &  4&1  & \\ 
\hline &1  &2  &3 \\ 
\hline  & 3 & 4 &1 \\ 
\hline 
\end{tabular}
\caption{The partially filled Latin Square obtained from $\mathcal{L}\left(-2\right).$}
\label{PFLS_2}
\end{figure}
\end{example}

The second step involves completion of the partially filled Latin Square obtained from the constrained partial Latin Square using the minimum number of symbols.
\begin{example}
Continuing with Example \ref{example7}, a completion of the partially filled Latin Square given in Fig. \ref{PFLS_2} using 4 symbols is provided in Fig. \ref{LS_2}. 
\begin{figure}
\centering
\begin{tabular}{|c|c|c|c|}
\hline  1&  2& 3 &4 \\ 
\hline 3&4  &1  &2 \\ 
\hline 4&1  &2  &3 \\ 
\hline 2 & 3 & 4 &1 \\ 
\hline 
\end{tabular}
\caption{A completion of the partially filled Latin Square given in Fig. \ref{PFLS_2}}
\label{LS_2}
\end{figure} 
\end{example} 

Now, we discuss the connection between the graph coloring approach and the two step procedure mentioned above. 
\begin{definition}
The vital subgraph of a singularity removal graph $\mathcal{G}_{s},$ denoted by $\mathcal{G}^V_{s},$ is the induced subgraph of $\mathcal{G}_{s}$ by the set of vertices $\{i : \vert c_i \vert \geq 2\}.$
\end{definition}

The process of obtaining a partially filled Latin Square with $k$ symbols from the constrained partial Latin Square, is equivalent to obtaining a proper $k$-coloring for the vital subgraph of the singularity removal graph. This is illustrated in the following examples. 
\begin{example}
\label{example9}
Continuing with Example \ref{example4}, the vital subgraph $\mathcal{G}^V_{\frac{(1+j)}{2}}$ of the singularity removal graph $\mathcal{G}_{\frac{(1+j)}{2}}$ is the complete graph with vertices 1, 2, 3 and 4, shown in Fig. \ref{fig:srg_critical1}. From Example \ref{example2}, it can be seen that the constraints $c_1,c_2,c_3$ and $c_4$ are of cardinality 2 and the other constraints are of cardinality 1. Note that for this case, the constrained partial Latin Square in Fig. \ref{CPLS_example} already has less than or equal to $M=4$ symbols and no more combining of symbols is required. Moreover, it can be verified that no more combining of symbols is possible. Equivalently, since the vital subgraph $\mathcal{G}^V_{\frac{(1+j)}{2}}$ is complete, 4 colors are needed for a proper coloring of its vertices, as shown in Fig. \ref{fig:srg_critical1}.
\begin{figure}[htbp]
\centering
\includegraphics[totalheight=.8in,width=3in]{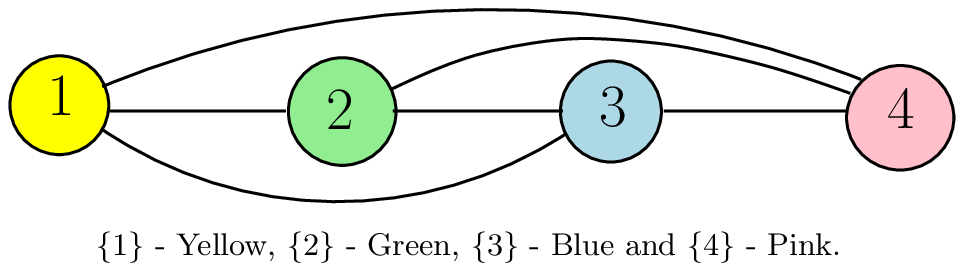}
\caption{Vital subgraph of the singularity removal graph $\mathcal{G}_{\frac{(1+j)}{2}}.$}
\label{fig:srg_critical1}
\end{figure}
\end{example}
\begin{example}
\label{example10}
Continuing with Example \ref{example7}, the singularity removal graph $\mathcal{G}_{-2}$ is as given in Fig. \ref{fig:srg_example2}. The vital subgraph $\mathcal{G}^V_{-2}$ is shown in Fig. \ref{fig:srg_critical2}. The partially filled Latin Square in Fig. \ref{PFLS_2} which has 4 symbols was obtained by combining the symbols 1 and 6, and also the symbols 3 and 5 in the constrained partial Latin Square given in Fig. \ref{CPLS_2}. Equivalently, the vital subgraph $\mathcal{G}^V_{-2}$ admits a proper 4-coloring with the vertices 1 and 6 colored using the same color yellow and the vertices 3 and 4 colored using the same color blue, as shown in Fig. \ref{fig:srg_critical2}. 
\begin{figure}[htbp]
\centering
\includegraphics[totalheight=2.2in,width=3in]{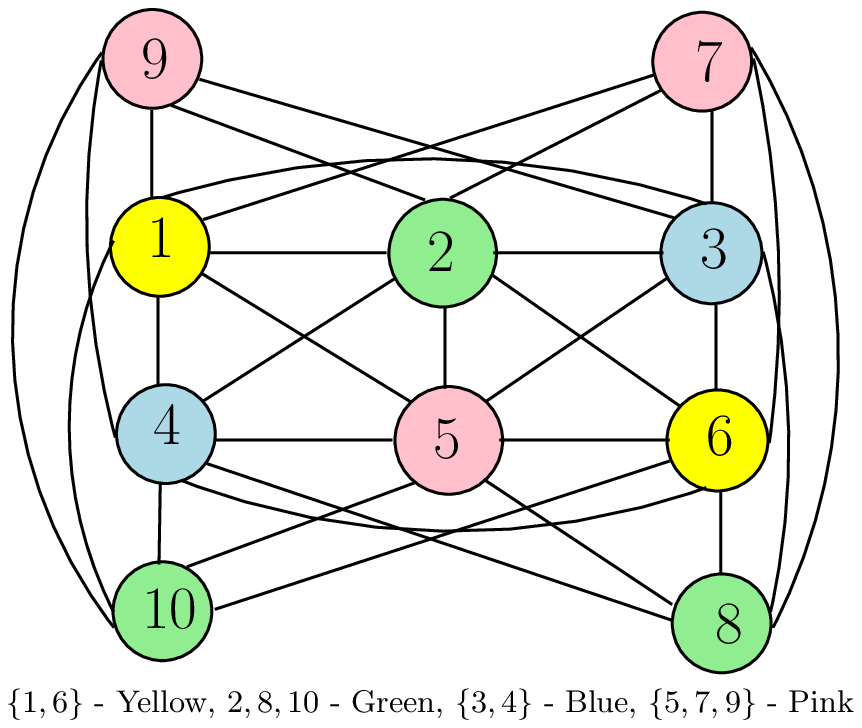}
\caption{The singularity removal graph $\mathcal{G}_{-2}.$}
\label{fig:srg_example2}
\end{figure}
\begin{figure}[htbp]
\centering
\includegraphics[totalheight=1.5in,width=3in]{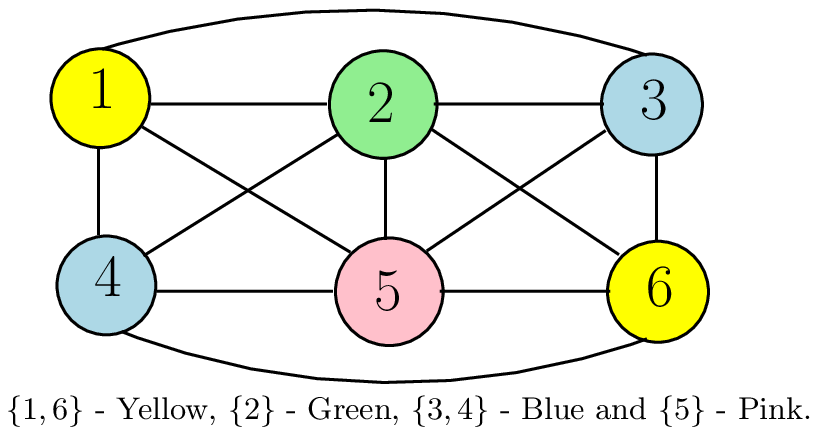}
\caption{Vital subgraph of the singularity removal graph $\mathcal{G}_{\frac{(1+j)}{2}}.$}
\label{fig:srg_critical2}
\end{figure}
\end{example}

The second step which involves completing the partially filled Latin Square can be viewed as coloring the rest of the singularity removal graph, whose subgraph, viz. the vital subgraph has already been colored. 
\begin{example}
Continuing with Example \ref{example9}, a proper 5-coloring can be obtained for the singularity removal graph $\mathcal{G}_{\frac{(1+j)}{2}}$ as shown in Fig. \ref{fig:srg_example}, with the coloring for the vertices which belong the vital subgraph $\mathcal{G}^V_{\frac{(1+j)}{2}}$ chosen to be the same as the one in Fig. \ref{fig:srg_critical1}.
\end{example}

\begin{example}
Continuing with Example \ref{example10}, a proper 4-coloring can be obtained for $\mathcal{G}_{-2}$ as shown in Fig. \ref{fig:srg_example2}, which was obtained by retaining the same coloring for the vertices which belong to the vital subgraph in Fig. \ref{fig:srg_critical2}.
\end{example}

To summarize, the two step procedure suggested in \cite{VNR} to obtain Latin Squares is equivalent to obtaining a proper vertex coloring for the singularity removal graph in two steps: first coloring the vital subgraph of the singularity removal graph to get a partially colored graph and then coloring the rest of the singularity removal graph. A disadvantage of this two step procedure is that the number of symbols in the completely filled Latin Square obtained after the second step depends on the way in which the symbols in the constrained partial Latin Square are combined in the first step. It is possible that due to the choice of combining symbols in the first step, the resulting Latin Square can use a larger number of symbols than the chromatic number of the singularity removal graph. This is illustrated in the following example.
\begin{example}
\label{example13}
Consider the case when 8-QAM signal set $\mathcal{S}=\{-3-j,-3+j,-1-j,-1+j,1-j,1+j,3-j,3+j\}$ is used at the end nodes, with the points labelled from 1 to 8 in the same order mentioned in the set $\mathcal{S}.$ It can be verified that $-0.5-0.5j$ is a singular fade state for this case and the corresponding constrained partial Latin Square $\mathcal{L}(-0.5-0.5j)$ is given in Fig. \ref{CPLS_3}. Combining the symbols (1 and 9), (2 and 15), (3 and 18), (4, 16 and 17), (5, 11 and 12), (6 and 13), (7 and 10), and (8 and 14) in $\mathcal{L}(-0.5-0.5j),$ one gets a partially filled Latin Square with 8 symbols shown in Fig. \ref{PFLS_3}. This partially filled Latin Square cannot be completed using 8 symbols since the (4,1)-th cell cannot be filled with any of the symbols 1 to 8. But this does not mean that there does not exist any Latin Square which removes the singular fade state $-0.5-0.5j$ which contains only 8 symbols. It can be verified that the Latin Square shown in Fig. \ref{LS_3} contains only 8 symbols, which was obtained by coloring the singularity removal graph $\mathcal{G}_{(-0.5-0.5j)}.$ 
\begin{figure}
\centering
\begin{tabular}{|c|c|c|c|c|c|c|c|}
\hline  1&  2& 3 &4& 5&6&& \\ 
\hline   &     7     & &     1&&     3 &  &     5 \\ 
\hline  8 &    9 &    2&    10&     4 &   11&     6& \\ 
\hline  &12&     7&     8&     1&     2&     3&     4 \\ 
\hline    13 &   14&     9&    15 &   10 &   16&    11 &\\
\hline  &   17 &   12 &   13 &    8 &    9&     2 &   10\\
\hline  18     &&    14     &&    15     &&    16     &\\
\hline  &     &    17&    18&    13 &   14 &    9&    15\\
\hline
\end{tabular}
\caption{The constrained partial Latin Square $\mathcal{L}(-0.5-0.5j).$}
\label{CPLS_3}
\end{figure}  
\begin{figure}
\centering
\begin{tabular}{|c|c|c|c|c|c|c|c|}
\hline  1&  2& 3 &4& 5&6&& \\ 
\hline   &     7     & &     1&&     3 &  &     5 \\ 
\hline  8 &    1 &    2&    7&     4 &   5&     6& \\ 
\hline  &5&     7&     8&     1&     2&     3&     4 \\ 
\hline    6 &   8&     1&    2 &   7 &   4&    5 &\\
\hline  &   4 &   5 &   6 &    8 &    1&     2 &   7\\
\hline  3     &&    8     &&    2     &&    4     &\\
\hline  &     &    4&    3&    6 &   8 &    1&    2\\
\hline
\end{tabular}
\caption{Partially filled Latin Square obtained from $\mathcal{L}(-0.5-0.5j).$}
\label{PFLS_3}
\end{figure}  
\begin{figure}
\centering
\begin{tabular}{|c|c|c|c|c|c|c|c|}
\hline  1&  2& 3 &4& 5&6&7&8 \\ 
\hline   4&     6     & 7&     1&2&     3 &8  &     5 \\ 
\hline  8 &    1 &    2&    7&     4 &   5&     6&3 \\ 
\hline  7&5&     6&     8&     1&     2&     3&     4 \\ 
\hline    3 &   8&     1&    6 &   7&   4&    5 &2\\
\hline  6&   4 &   5 &   3 &    8 &    1&     2 &   7\\
\hline  2     &3&    8     &5&    6     &7&    4     &1\\
\hline  5&  7   &    4&    2&    3 &   8 &    1&    6\\
\hline
\end{tabular}
\caption{A Latin Square which removes the singular fade state $-0.5-0.5j.$}
\label{LS_3}
\end{figure}  
\end{example}

It was illustrated in Example \ref{example13} that the number of symbols in the Latin Square obtained after the second step depends on the way in which the symbols in the constrained partial Latin Square are combined in the first step. Note that there always exists a way of combining symbols in the first step which would result in a Latin Square with the minimum number of symbols. This follows from the fact that going backwards, such a way of combining symbols can be obtained from a Latin Square which contains the minimum number of symbols. But while carrying out the first step, in general, it is not known which way of combining constraints will result in Latin Square with the minimum number of symbols. In Section V, for $2^{\lambda}$-PSK signal set, a specific way of combining symbols in the constrained partial Latin Square is presented, which ensures that the Latin Square obtained contains the minimum number of symbols. 
\section{Removal of Singular Fade States with Latin Squares containing $M$ Symbols}
To satisfy the condition that no two entries in a row (column) can be the same, a Latin Square should contain at least $M$ symbols. Due to the additional constraints imposed by singular fade state removal, there may not exist a Latin Square with only $M$ symbols which removes a singular fade state. As mentioned earlier in Section III-A, the minimum number of symbols required in a Latin Square used for the removal of a singular fade state $s$ is equal to the chromatic number $\chi(\mathcal{G}_s)$ of the singularity removal graph $\mathcal{G}_s.$ If $\chi(\mathcal{G}_s)> M,$ there does not exist a Latin Square with $M$ symbols which removes $s.$ Since $\chi(\mathcal{G}_s)$ is lower bounded by the clique number $\omega(\mathcal{G}_s),$ a sufficient condition under which $\chi(\mathcal{G}_s)> M$ is the existence of a clique of size greater than $M.$
\begin{example}
Continuing with Example \ref{example4}, it can be seen that the subgraph of $\mathcal{G}_{\frac{(1+j)}{2}}$ in Fig. \ref{fig:srg_example}, induced by the vertices 1, 2, 3, 6 and 7 form a clique. Hence, we have $\chi(\mathcal{G}_{\frac{(1+j)}{2}})\geq 5.$ In fact, $\chi(\mathcal{G}_{\frac{(1+j)}{2}})= 5$ since a proper 5-coloring has been already shown in Fig. \ref{fig:srg_example}. 
\end{example} 

 It can be verified that for 4-QAM signal set $\chi(\mathcal{G}_{s})= 5$ when $s$ belongs to the set $\{\pm1\pm j, \pm 0.5\pm 0.5j\}.$ In other words, for 4-QAM signal set there exists at least 8 singular fade states, which can not be removed by Latin Squares which contain the minimum number of symbols $M.$ In the following theorem, it is shown that this is valid for any square $M$-QAM signal set.
 \begin{theorem}
 \label{thm1}
 For any square $M$-QAM signal set, we have $\chi(\mathcal{G}_{s})\geq\omega(\mathcal{G}_s)\geq M+1,$ for $s \in  \{\pm1\pm j, \pm 0.5\pm 0.5j\}.$ Hence, for any square QAM signal set, there exists at least 8 singular fade states which can not be removed by Latin Squares which contain the minimum number of symbols $M.$ 
 \begin{proof}
 The points in the square $M$-QAM signal set $\mathcal{S}$ are of the form $(-\sqrt{M}+1+2l)+(-\sqrt{M}+1+2k)j,$ where $k,l \in \{0,1,\dotso,\sqrt{M}-1\}.$ Let the points be indexed from 1 to $M,$ with the point $(-\sqrt{M}+1+2l)-(-\sqrt{M}+1+2k)j$ indexed by $k+l\sqrt{M}+1.$ Note that $\pm 1\pm j$ and $\pm 0.5\pm 0.5j$ are singular fade states for any $M$- QAM signal set. The reason is as follows: the singular fade states are of the form $-(x_A-x'_A)/(x_B-x'_B),x_A,x_B,x'_A,x'_B\in \mathcal{S}.$ Since difference between points in $\mathcal{S}$ can take values $\pm 2\pm 2j$ and $\pm 2,$ the complex numbers $\pm 1\pm j$ and $\pm 0.5\pm 0.5j$ are singular fade states.

It is enough to prove the result $s=-1-j.$ The proof for the other 7 values of $s$ follows from the fact that the Latin Squares which these 7 singular fade states can be obtained from the one which removes $s=-1-j$ by appropriate column permutation or by taking transpose \cite{NVR}. 

Consider the set of singularity removal constraints $\mathcal{C}({-1-j}).$ Note that every element of $\mathcal{S} \times \mathcal{S}$ belongs to exactly one of the singularity removal constraints. Let $c_i, i \in \{1,2,\dotso, M-\sqrt{M}\}$ denote the singularity removal constraint which contains $(1,i).$ Let $i=k+l\sqrt{M}+1,$ where $k,l \in \{0,1,\dotso,\sqrt{M}-1\}.$

Let $x_A$ and $x_B$ denote the signal points labelled by 1 and $i$ respectively, which are given by $(-\sqrt{M}+1)+(-\sqrt{M}+1)j$ and $(-\sqrt{M}+1+2l)+(-\sqrt{M}+1+2k)j$ respectively. Let $x'_A$ denote the signal point labelled by $\sqrt{M}+2$ which is given by $(-\sqrt{M}+3)+(-\sqrt{M}+3)j.$ For $i \in \{1,2,\dotso, M-\sqrt{M}\},$ let $x'_B$ denote the signal point labelled by $i+\sqrt{M}$ which is given by $(-\sqrt{M}+3+2l)+(-\sqrt{M}+1+2k)j.$ It can be verified that $x_A+(-1-j)x_B=x'_A+(-1-j)x'_B.$ Hence, the singularity removal constraint $c_i$ also contains $(\sqrt{M}+2,\sqrt{M}+i)$ in addition to $(1,i).$  

For $i \in \{M-\sqrt{M}+1,M-\sqrt{M}+2,\dotso,M\},$ let $c_i$ denote the singularity removal constraint which contains $(\sqrt{M}+2,i-M+\sqrt{M}).$ The $M$ singularity removal constraints $c_i, i \in \{1,2,\dotso,M\}$ contain at least one element of the form $(a,b_i)$ where $a=\sqrt{M}+2.$ Hence, from the definition of a singularity removal graph, it follows that for the singularity removal graph $\mathcal{G}_{-1-j},$ the induced subgraph formed by the nodes $\{1,2,\dotso,M\}$ forms a clique of size $M.$ To complete the proof of the theorem, it suffices to show that this clique is contained in a larger clique which contains one more node.    

Let $c_{M+1}$ denote a singularity removal constraint which contains $(2,\frac{(M-\sqrt{M}+2)}{2}).$ It is claimed that $(2,\frac{(M-\sqrt{M}+2)}{2})$ does not belong to any of the $M$ singularity removal constraints $c_i, i \in \{1,2,\dotso,M\}$ defined earlier. The proof of this claim is follows: For $i \in \{1,2,\dotso,M\},$ the singularity removal constraint $c_i$ contains an element of the form $(\sqrt{M}+2,b_i).$ The signal point labelled by $\sqrt{M}+2$ is $(-\sqrt{M}+3)+(-\sqrt{M}+3)j.$ Let $x(b_i)$ denote the signal point labelled by $b_i.$ Note that the points labelled by $2$ and $\frac{(M-\sqrt{M}+2)}{2}$ are respectively $(-\sqrt{M}+1)+(-\sqrt{M}+3)j$ and $-1+j.$ Hence, for all $(x_A,x_B)\in c_{M+1},$ we have $x_A+(-1-j)x_B=(-\sqrt{M}+3)+(-\sqrt{M}+3)j.$ If $(2,\frac{(M-\sqrt{M}+2)}{2})$ belongs to any of the $M$ singularity removal constraints $c_i, i \in \{1,2,\dotso,M\},$ then the following condition should hold: $((-\sqrt{M}+3)+(-\sqrt{M}+3)j)+(-1-j)x(b_i)=(-\sqrt{M}+3)+(-\sqrt{M}+3)j.$ The above condition cannot hold since $x(b_i)=0$ is not a valid point in the square $M$-QAM signal set. 

Let the singularity removal constraints $c_i, i \in \{1,2,\dotso, M\},$ be classified as $c_{u+v\sqrt{M}+1},$ where $u \in \{0,1,2,\dotso,\sqrt{M}-1\}$ and $v \in \{0,1,\dotso, \sqrt{M}-1\}.$

We claim that there exits $(a_{u,v},b_u)$ and $(a'_u,b_u)$ such that $(a_{u,v},b_u)\in c_{u+v\sqrt{M}+1}$ and $(a'_u,b_u) \in c_{M+1}.$ To complete the proof of this theorem, it suffices to prove this claim.

First it is shown that for $b_u=M-2\sqrt{M}+u+1, u \in \{0,1,\dotso,\sqrt{M}-1\},$ there exists $(a'_u,b_u)\in c_{M+1}.$ As mentioned earlier, for all $(x_A,x_B)\in c_{M+1},$ we have $x_A+(-1-j)x_B=(-\sqrt{M}+3)+(-\sqrt{M}+3)j.$ Note that the point labelled as $b_u$ is $(\sqrt{M}-3)+(-\sqrt{M}+2u+1)j.$ To show that $(a'_u,b_u)\in c_{M+1},$ we need to show the existence of a signal point $x(a'_u)$ which is labelled by $a'_u,$ which satisfies $x(a'_u)+(-1-j)((\sqrt{M}-3)+(-\sqrt{M}+2u+1)j)= (-\sqrt{M}+3)+(-\sqrt{M}+3)j.$ Solving for $x(a'_u),$ we get $x(a'_u)=(\sqrt{M}-2u-1)+(-\sqrt{M}+2u+1)j,$ which is indeed a valid signal point for $u\in \{0,1,\dotso,\sqrt{M}-1\}.$

To complete the proof, it suffices to show that there exists $(a_{u,v},M-2\sqrt{M}+u)\in c_{u+v\sqrt{M}+1}.$ First we consider $u \in \{0,1,\dotso \sqrt{M}-1\}$ and $v \in \{0,1\dotso,\sqrt{M}-2\},$ for which $u+v\sqrt{M}+1\in\{1,2,\dotso, M-\sqrt{M}\}.$ Note that $(1,u+v\sqrt{M}+1)\in c_{u+v\sqrt{M}+1},$ and the labels 1 and $u+v\sqrt{M}+1$ respectively denote the signal points $-\sqrt{M}+1+(-\sqrt{M}+1)j$ and $(-\sqrt{M}+2v+1)+(-\sqrt{M}+2u+1)j.$ Hence, every $(x_A,x_B)\in c_{u+v\sqrt{M}+1}$ satisfies $x_A+(-1- j)x_B=(-\sqrt{M}+1+2u-2v)+j(\sqrt{M}-1-2u-2v).$ To show the existence of $(a_{u,v},b_u)\in c_{u+v\sqrt{M}+1},$ it suffices to show the existence of $x(a_{u,v}) \in \mathcal{S}$ which satisfies $x(a_{u,v})+(-1-j)((\sqrt{M}-3)+(-\sqrt{M}+2u+1)j)=(-\sqrt{M}+1+2u-2v)+j(\sqrt{M}-1-2u-2v).$ Solving for $x(a_{u,v}),$ we get $(\sqrt{M}-2v-3)+(\sqrt{M}-2v-3)j,$ which indeed belongs to $\mathcal{S}$ for $v \in \{0,1,2,\dotso,\sqrt{M}-2\}.$

Finally, to complete the proof, the existence of $(a_{u,v},b_u)\in c_{u+v\sqrt{M}+1},$ needs to be shown for $v=\sqrt{M}-1$ and $u \in \{0,1,\dotso,\sqrt{M}-1\},$ for which $u+v\sqrt{M}+1\in \{M-\sqrt{M}+1,M-\sqrt{M}+2,\dotso M\}.$ Since, for this case, $(\sqrt{M}+2,u+1)\in c_{u+v\sqrt{M}+1},$ every $(x_A,x_B)\in c_{u+v\sqrt{M}+1}$ satisfies $x_A+(-1-j)x_B=(-\sqrt{M}+3+2u)+(\sqrt{M}+1-2u)j.$ We need to find $x(a_{u,v})\in \mathcal{S}$ which satisfies $x(a_{u,v})+(-1-j)((\sqrt{M}-3)+(-\sqrt{M}+2u+1)j)=(-\sqrt{M}+3+2u)+(\sqrt{M}+1-2u)j.$ Solving, we get $x(a_{u,v})=(\sqrt{M}-1)+(\sqrt{M}-1)j,$ which belongs to $\mathcal{S}.$ This completes the proof of Theorem \ref{thm1}. 
 \end{proof}
 \end{theorem}
 \section{Singular Fade State Removal using Latin Squares with $M$ symbols for $M$-PSK Signal Set}
In the previous section, it was shown that for square $M$-QAM signal set, there exists singular fade states which cannot be removed using Latin Squares containing only $M$ symbols. In this section, we prove the following result which was conjectured in \cite{VNR}: for any $M$-PSK signal set ($M$ of the form $2^{\lambda}$ and $M\geq 8$), all the singular fade states can be removed using Latin Squares containing only $M$ symbols.

A brief outline of the results in \cite{VNR} related to singularity removal for $2^{\lambda}$-PSK signal set are presented in Section V-A. To show the existence of Latin Squares with $M$ symbols, the two step procedure explained in Section III B is adopted. First, in Section V-B, an explicit coloring of the vital subgraph of the singularity removal graph is provided, which would result in a partially filled Latin Square. In Section V-C, it is shown that the partially filled Latin Squares thus obtained are completable using $M$ symbols.
\subsection{Singular fade states and Constrained Partial Latin Squares for $2^{\lambda}$-PSK signal set}
In this subsection, a brief outline of the results in \cite{VNR} related to singularity removal for $2^{\lambda}$-PSK signal set is presented. For more details and proofs, refer \cite{VNR}. Consider the $M$-PSK signal set $\{e^{j(2m-1)\pi /M}:m \in \{1,2,\dotso,M\}\},$ with the signal point $e^{j(2m-1)\pi /M}$ labelled as $m,$ where $M \geq 8.$  
For $M$-PSK signal set, $M$ of the form $2^{\lambda},$ there are $(M^{2}/4-M/2+1)M$ singular fade states. These singular fade states lie on $M^2/4-M/2+1$ circles with radii of the form $\frac{\sin(k\pi/M)}{\sin(l\pi/M)},$ $k ,l \in \{1,2,\dotso,M/2\}$ with $M$ points lying on each circle. The $M$-points on a circle with radius $\frac{\sin(k\pi/M)}{\sin(l\pi/M)}$ have phase angles 
\begin{itemize}
\item[(i)]
${2m\pi/M},m\in\{0,1,\dotso M-1\}$ if both $k$ and $l$ are odd or both are even, 
\item[(ii)]
 ${(2m+1)\pi/M},m\in\{0,1,\dotso M-1\}$ if only one among $k$ and $l$ is even. 
\end{itemize} 
 
 From a Latin Square which removes a singular fade state which lies on a circle of radius $r,$ Latin Squares which remove the other singular fade states with radius $r$ can be obtained by appropriate column permutation \cite{VNR}. Also, the conventional bit-wise exclusive OR Latin Square removes the singular fade state $s=1$ \cite{VNR}. Hence, in the rest of the paper, we consider only the singular fade states of the form 
 \begin{itemize}
 \item[(i)]
 $\frac{\sin(\pi k/M)}{\sin(\pi l/M)},$ when both $k$ and $l$ are odd or both are even,  
 \item[(ii)]
  $\frac{\sin(\pi k/M)}{\sin(\pi l/M)}e^{j \pi/M}$ when only one among $k$ and $l$ is odd, 
  \end{itemize}  
  where $k,l\in\{1,2,\dotso M/2\}$ and $k \neq l.$



For a singular fade state of the form $\frac{\sin(\pi k/M)}{\sin(\pi l/M)},$ where both $k$ and $l$ are even or both are odd, the singularity removal constraints  with cardinality greater than two are given by \cite{VNR},

{\scriptsize
\begin{align}
\nonumber
&\left\lbrace\left(i+1,\left(i-\frac{M}{2}-\frac{k-l}{2}\right) \text{mod}\: M +1\right)\right.,\\
\label{eqn1}
&\underbrace{\hspace{1.25 cm}\left.\left(\left(i-k\right)\text{mod}\:M +1,\left(i+\frac{M}{2}-\frac{k+l}{2}\right)\text{mod}\: M +1\right)\right\rbrace}_{c_{i+1}},\\
\nonumber
&\left\lbrace\left(i+1,\left(i-\frac{k+l}{2}\right)\text{mod}\: M +1\right),\right.\\
\label{eqn2}
&\underbrace{\hspace{1.5 cm}\left.\left(\left(i-k\right)\text{mod}\: M+1,\left(i-\frac{k-l}{2}\right)\text{mod}\: M+1\right)\right\rbrace}_{c_{M+i+1}},
\end{align}
}where $0 \leq i \leq M-1$. For this case, for $i \in \{0,2,\dotso, M-1\},$ let $c_{i+1}$ and $c_{i+M+1}$ respectively denote the singularity removal constraints as in \eqref{eqn1} and \eqref{eqn2}. 

For the singular fade state $\frac{\sin(\pi k/M)}{\sin(\pi l/M)}e^{j\pi/M},$ where only one among $k$ and $l$ is odd, the singularity removal constraints  with cardinality greater than two are given by,

{\scriptsize
\begin{align}
\nonumber
&\left\lbrace\left(i+1,\left(i-\frac{M}{2}-\frac{k+1-l}{2}\right)\text{mod}\:M+1\right),\right.\\
\label{eqn3}
&\underbrace{\hspace{0.75 cm}\left.\left(\left(i-k\right)\text{mod}\:M+1,\left(i+\frac{M}{2}-\frac{k+1+l}{2}\right)\text{mod}\:M+1\right)\right\rbrace}_{c_{i+1}},\\
\nonumber
&\left\lbrace\left(i+1,\left(i-\frac{\left(k+1+l\right)}{2}\right)\text{mod}\:M+1\right),\right.\\
\label{eqn4}
&\underbrace{\hspace{1.05 cm}\left.\left(\left(i-k\right)\text{mod}\:M+1,\left(i-\frac{\left(k+1-l\right)}{2}\right)\text{mod}\:M+1\right)\right\rbrace}_{c_{M+i+1}},
\end{align}
}where $0 \leq i \leq M-1$.  For this case, for $i \in \{0,2,\dotso, M-1\},$ let $c_{i+1}$ and $c_{i+M+1}$ respectively denote the singularity removal constraints as in \eqref{eqn3} and \eqref{eqn4}. 

For the case when $k\neq M/2$ and $l \neq M/2,$ the $2M$ constraints given in \eqref{eqn1} and \eqref{eqn2} (and also \eqref{eqn3} and \eqref{eqn4}) are distinct. Hence, the constrained partial Latin Square for this case contains $2M$ symbols, with each symbol appearing twice and four cells filled in every row and every column.

For the case when $k=M/2$ or $l=M/2,$ it can be seen from \eqref{eqn1} and \eqref{eqn2} (and also \eqref{eqn3} and \eqref{eqn4}) that only $M$ constraints, $c_i, i \in \{1,2\dotso M\},$ are distinct. Hence, the constrained partial Latin Square for this case is filled in with $M$ symbols, with each symbol appearing twice and two cells filled in every row and every column. 
\subsection{Vertex Coloring of the vital subgraphs to obtain Partially filled Latin Squares}  
In this subsection, depending on the singular fade state, a proper 4-coloring or 8-coloring is obtained for the vital subgraphs of the singularity removal graphs.  Note that the number of colors used to color the vital subgraphs of the singularity removal graphs is either 4 or 8, irrespective of the size of the signal set $M.$   

First, the vital subgraphs of the singularity removal graphs are characterized for the different singular fade states. 

The vital subgraph of the singularity removal graph for a singular fade state of the form $\frac{\sin(\pi k /M)}{\sin(\pi l/M)},$ where $k\neq M/2,l\neq M/2,$ is given by the following lemma.
\begin{lemma}
\label{lemma1}
For a singular fade state $s$ of the form (i) $\frac{\sin(\pi k /M)}{\sin(\pi l/M)},$ when both $k$ and $l$ are odd or both are even, or (ii) $\frac{\sin(\pi k /M)}{\sin(\pi l/M)}e^{j \pi/M},$ when only one among $k$ and $l$ is odd, where $k\neq M/2,l\neq M/2,$ the vital subgraph $\mathcal{G}^V_s$ is as follows: $\mathcal{G}^V_s$ has vertex set $\{1,2,\dotso,2M\}.$ For $i\in \{0,1,\dotso M-1\},$ 
\begin{itemize}
\item
 vertex $i+1$ is adjacent to vertices in the set 

{\scriptsize
\begin{align*}
& \left\{\left(i\pm k\right)\text{mod}\: M+1,\left(i\pm l\right) \text{mod}\: M+1,M+i+1,\right.\\
&\hspace{1.2cm}M+\left(\left(i \pm k\right)\text{mod}\: M\right)+1,\left. M+\left(\left(M/2+i \pm l\right)\text{mod}\: M\right)+1,\right.\\
&\hspace{4.8cm}\left. M+\left(i + M/2\right)\text{mod}\: M+1\right\}.
\end{align*} 
 }
 
\item 
  vertex $M+i+1$ is adjacent to vertices in the set 

{\scriptsize
\begin{align*}
& \left\{i+1,\left(i\pm k\right)\text{mod}\: M+1,\left(M/2+i\pm l\right) \text{mod}\: M+1,\right.\\
&\hspace{1.2 cm}M+\left(\left(i \pm k\right)\text{mod}\: M\right)+1,\left. M+\left(\left(i \pm l\right)\text{mod}\: M\right)+1,\right.\\
&\hspace{4.8 cm}\left.\left(i + M/2\right)\text{mod}\: M+1\right\}.
\end{align*} 
 }
\end{itemize}
\begin{proof}
The proof is given for the case when $s$ is of the form $\frac{\sin(\pi k /M)}{\sin(\pi l/M)},$ where both $k$ and $l$ are odd or both are even. The proof for the case when only one among $k$ and $l$ is odd is similar and is omitted.
$\mathcal{G}^V_s$ has vertex set $\{1,2,\dotso,2M\},$ since there are $2M$ singularity removal constraints, which are given in \eqref{eqn1} and \eqref{eqn2}. To show that vertex $i$ is adjacent to vertex $j,$ it needs to be shown that there exits $(a,b)\in c_i$ and $(a',b')\in c_j$ such that $a=a'$ or $b=b'.$  Vertex $i+1$ is adjacent to vertex $\left(i-k\right)\text{mod}\:M+1$ since \mbox{\scriptsize$\left(\left(i-k\right)\text{mod}\:M+1,\left(i+\frac{M}{2}-\frac{\left(k+l\right)}{2}\right)\text{mod}\:M+1\right)\in c_{i+1}$} and \mbox{\scriptsize$ \left(\left(i-k\right)\text{mod}\:M+1,\left(i-\frac{M}{2}-\frac{\left(3k-l\right)}{2}\right)\text{mod}M+1\right)\in c_{(i-k)\text{mod}\:M+1}.$} Similarly, it can be shown that the vertex $i+1$ is adjacent to the other vertices given in the statement of the lemma. By a similar reasoning the vertex $M+i+1$ is adjacent to the set of vertices given in the statement of the lemma.
\end{proof}
\end{lemma}

%
%
%

For a singular fade state with absolute value $\sin(\pi k/M)$   or $\frac{1}{\sin(\pi k /M)},$ $\mathcal{G}^V_s$ is provided in the following lemma.
\begin{lemma}
\label{lemma3}
For singular fade states of the form  (i) $s={\sin(\pi k /M)}\:\text{or}\:\frac{1}{\sin(\pi k/M)},$ when $k$ is  even or (ii) $s=\sin(\pi k /M)e^{j\pi/M}\:\text{or}\:\frac{1}{\sin(\pi k/M)}e^{j \pi/M}$ when $k$ is odd, $\mathcal{G}^V_s$ is as follows: $\mathcal{G}^V_s$ has vertex set $\{1,2,\dotso,M\}.$ For $i\in \{0,1,\dotso M-1\},$ vertex $i+1$ is adjacent to the vertices in the set $\{(i\pm k)\text{mod}\:M+1,(i+M/2)\text{mod}\:M+1\}.$
\begin{proof}
The proof is similar to that of Lemma \ref{lemma1} and is omitted.
\end{proof}
\end{lemma}

Now we proceed to obtain proper vertex colorings for the different vital subgraphs of the singularity removal graphs obtained above. Note that in the vital subgraph obtained in Lemma \ref{lemma1}, two vertices $i$ and $j,$ both of which are either less than or equal to $M$ or both are greater than $M,$ are adjacent only if they differ by either $k$ or $l$ in modulo-$M$ arithmetic. 
In the proper vertex colorings provided below, the colors used for the vertices which are less than or equal to $M$ are chosen to be different from the colors used for vertices greater than $M.$ To ensure that the colorings provided are proper, two vertices which differ by either $k$ or $l$ in modulo-$M$ arithmetic are assigned different colors. 

For a singular fade state of the form $\frac{\sin(\pi k/M)}{\sin(\pi l /M)},$ both $k$ and $l$ being odd, there exists a proper 4-coloring for $\mathcal{G}^V_s$, as shown in the following lemma.
\begin{lemma}
\label{lemma4}
For a singular fade state $s=\frac{\sin(\pi k/M)}{\sin(\pi l /M)},$ both $k$ and $l$ being odd, $\mathcal{G}^V_s$ has a proper 4-coloring in which the  vertices which belong to the four sets $\{1,3,\dotso,M-1\},\{2,4,\dotso,M\},\{M+1,M+3,\dotso, 2M-1\},\{M+2,M+4,\dotso, 2M\},$ are assigned four different colors.
\begin{proof}
For $i\leq M$ and $j\leq M,$ from Lemma \ref{lemma1} it follows that a vertex $i,$ where $i$ is odd (even), is adjacent to vertex $j$ only if $j$ is even (odd), since $k$ and $l$ are odd. Hence, the vertices which belong to the set $\{1,3,\dotso,M-1\},$ ($\{2,4,\dotso,M\}$) are non-adjacent and can be colored using the same color. Similarly, for $i\leq M$ and $j\leq M,$ a vertex $M+i,$ i being odd (even), is adjacent to a vertex $M+j$ only if $j$ is even (odd). Hence, the vertices which belong to the set $\{M+1,M+3,\dotso, 2M-1\},$ ($\{M+2,M+4,\dotso, 2M\}$) are non-adjacent and can be colored using the same color.
\end{proof}
\end{lemma}

For $i\in\{1,2,\dotso,2^{\lambda-m}\},$ let $U_i$ and $V_i$ respectively denote the sets $\{(i-1)2^{m}+1,(i-1)2^{m}+2,\dotso,i 2^{m}\}$ and $\{M+(i-1)2^{m}+1,M+(i-1)2^{m}+2,\dotso,M+i 2^{m}\}.$ 

For $m_1<m_2,$ $i \in \{1,2,\dotso,2^{m_2-m_1+1}\}$ and $j\in \{1,2,\dotso,2^{\lambda-m_2-1}\},$ let $U_{i,j}$ and $V_{i,j}$ denote the sets,

{\scriptsize
\begin{align*}
&U_{i,j}=\{(j-1)(2^{m_2+1})+(i-1)2^{m_1}+1,\\
&\hspace{1cm}(j-1)(2^{m_2+1})+(i-1)2^{m_1}+2,\dotso,(j-1)(2^{m_2+1})+i 2^{m_1}\},\\
&V_{i,j}=\{(j-1)(2^{m_2+1})+(i-1)2^{m_1}+1+M,\\
&\hspace{0cm}(j-1)(2^{m_2+1})+(i-1)2^{m_1}+2+M,\dotso,(j-1)(2^{m_2+1})+i 2^{m_1}+M\}.
\end{align*}
}

Lemma \ref{lemma5} and Lemma \ref{lemma6} below deal with proper vertex colorings for singular fade states of the form $\frac{\sin(\pi k/M)}{\sin(\pi l /M)},$ where $k$ and $l$ are even, and $k\neq M/2,l\neq M/2.$
\begin{lemma}
\label{lemma5}
Consider a singular fade state $s=\frac{\sin(\pi k/M)}{\sin(\pi l /M)},$ where $k$ and $l$ are of the form $k=2^{m}a_1$ and $l=2^{m}a_2,$  $a_1$ and $a_2$ being odd integers and, $k\neq M/2,l\neq M/2.$ For this case, $\mathcal{G}^V(s)$ has a proper 4-coloring as given below.  The  vertices which belong to the following 4 sets are assigned 4 different colors:
\begin{align*}
{\bigcup_{\substack{{1\leq i\leq 2^{\lambda-m}}\\ {i \: \text{odd}}}}}\hspace{-0.35 cm}U_i,\hspace{.2 cm}
{\bigcup_{\substack{{1\leq i\leq 2^{\lambda-m}}\\ {i \: \text{even}}}}}\hspace{-0.35 cm}U_i, \hspace{.2 cm}
{\bigcup_{\substack{{1\leq i\leq 2^{\lambda-m}}\\{ i \: \text{odd}}}}}\hspace{-0.35 cm}V_i, \hspace{.2 cm}
{\bigcup_{\substack{{1\leq i\leq 2^{\lambda-m}}\\{ i \: \text{even}}}}}\hspace{-0.35 cm}V_i.
\end{align*}
\begin{proof}
Since $k$ and $l$ are odd multiples of $2^{m},$ from Lemma \ref{lemma1}, it can be seen that a vertex $i\leq M$ is adjacent to vertex $j\leq M$ only if they differ by an odd multiple of $2^{m}$ in modulo $M$ arithmetic. Since no two elements of the set $\displaystyle{{\bigcup_{\substack{{1\leq i\leq 2^{\lambda-m}}\\ {i \: \text{odd}}}}}\hspace{-0.35 cm}U_i,}$ differ by an odd multiple of $2^m$ in modulo $M$ arithmetic, they can be colored using the same color. By a similar reasoning, it follows that the vertices which belong to one among the other three sets given in the statement of the lemma can be colored using the same color. 
\end{proof}
\end{lemma}
\begin{lemma}
\label{lemma6}
Consider a singular fade state $s=\frac{\sin(\pi k/M)}{\sin(\pi l /M)},$ where $k$ and $l$ are of the form $k=2^{m_1}a_1$ and $l=2^{m_2}a_2,$  $a_1,a_2$ being odd integers and $m_1 <m_2.$ 
For this case, $\mathcal{G}^V(s)$ has a proper 8-coloring in which the vertices belonging to the following 8 sets are assigned distinct colors:

{\scriptsize
\begin{align*}
&{\bigcup_{\substack{{1\leq i\leq 2^{m_2-m_1},}\\{i \:\text{odd},}\\{1\leq j\leq 2^{\lambda-m_2-1}}}} \hspace{-.5 cm}U_{i,j}},
{\bigcup_{\substack{{1\leq i\leq 2^{m_2-m_1},}\\{i \:\text{even},}\\{1\leq j\leq 2^{\lambda-m_2-1}}}} \hspace{-.5 cm}U_{i,j}},
{\bigcup_{\substack{{2^{m_2-m_1}+1\leq i\leq 2^{m_2-m_1+1},}\\{i \:\text{odd}}\\{1\leq j\leq 2^{\lambda-m_2-1}}}} \hspace{-.5 cm}U_{i,j}},\\
&{\bigcup_{\substack{{2^{m_2-m_1}+1\leq i\leq 2^{m_2-m_1+1},}\\{i \:\text{even}}\\{1\leq j\leq 2^{\lambda-m_2-1}}}} \hspace{-0.5 cm} U_{i,j}},
{\bigcup_{\substack{{1\leq i\leq 2^{m_2-m_1},}\\{i \:\text{odd}}\\{1\leq j\leq 2^{\lambda-m_2-1}}}} \hspace{-.5 cm}V_{i,j}},
{\bigcup_{\substack{{1\leq i\leq 2^{m_2-m_1},}\\{i \:\text{even}}\\{1\leq j\leq 2^{\lambda-m_2-1}}}}\hspace{-.5 cm} V_{i,j}},\\
&{\bigcup_{\substack{{2^{m_2-m_1}+1\leq i\leq 2^{m_2-m_1+1},}\\{i \:\text{odd}}\\{1\leq j\leq 2^{\lambda-m_2-1}}}} \hspace{-.5 cm}V_{i,j}},
{\bigcup_{\substack{{2^{m_2-m_1}+1\leq i\leq 2^{m_2-m_1+1},}\\{i \:\text{even}}\\{1\leq j\leq 2^{\lambda-m_2-1}}}} \hspace{-.5 cm} V_{i,j}}.
\end{align*}}
\begin{proof}
For $i\leq M$ and $j\leq M,$ from Lemma \ref{lemma1}, it follows that the two vertices $i$ and $j$ can be adjacent only if they differ by an odd multiple of $2^{m_1}$ or $2^{m_2}$ in modulo $M$ arithmetic. Since no two elements of the union of the sets $U_{i,j},$ $i \in \{1,3,\dotso, 2^{m_2-m_1}-1\},$ $j \in \{1,2,\dotso,2^{\lambda-m_2-1}\},$  differ by any odd multiple of $2^{m_1}$ or $2^{m_2}$ in modulo $M$ arithmetic, they can be colored using the same color. Similarly, it follows that the vertices which belong to one among the other seven sets given in the statement of the lemma can be colored using the same color. 
\end{proof}
\end{lemma}
\begin{remark}
In Lemma \ref{lemma6}, only those singular fade states for which $m_1<m_2$ have been considered. If $m_1>m_2,$ Lemma \ref{lemma6} can be applied for the singular fade state  $s=\frac{\sin(\pi l/M)}{\sin(\pi k /M)}.$ Note that a Latin Square which removes the singular fade state $\frac{\sin(\pi k/M)}{\sin(\pi l /M)}$ can be obtained from a Latin Square which removes $\frac{\sin(\pi l/M)}{\sin(\pi k /M)},$ by taking transpose \cite{VNR}.
\end{remark}

For $i \in \{1,2,\dotso,2^{\lambda-m}\},$ let $U^{odd}_i$ and $U^{even}_i$ respectively denote the sets $\{1+(i-1)2^{m},3+(i-1)2^{m},\dotso,i2^{m}-1\}$ and $\{2+(i-1)2^{m},4+(i-1)2^{m},\dotso,i2^{m}\}.$ Also, for $i \in \{1,2,\dotso,2^{\lambda-m}\},$ let $V^{odd}_i$ and $V^{even}_i$ respectively denote the sets $\{M+1+(i-1)2^{m},M+3+(i-1)2^{m},\dotso,M+i2^{m}-1\}$ and $\{M+2+(i-1)2^{m},M+4+(i-1)2^{m},\dotso,M+i2^{m}\}.$

The following lemma provides a proper 8 coloring for those singular fade states of the form $\frac{\sin(k\pi/M)}{\sin(l\pi/M)}e^{j \pi/M},$ where one among $k$ and $l$ is odd and the other is an even integer other than $M/2.$
\begin{lemma}
\label{lemma7}
Consider a singular fade state $s=\frac{\sin(k\pi/M)}{\sin(l\pi/M)}e^{j \pi/M},$ where one among $k$ and $l$ is odd and the other is of the form $2^{m}a_1,$ $a_1$ being odd, and, $k\neq M/2,l\neq M/2.$  To obtain a proper 8-coloring for $\mathcal{G}_s^C,$ assign 8 different colors to the vertices which belong to the following 8 sets: 

{\footnotesize
\begin{align*}
&{\bigcup_{\substack{{1 \leq i \leq 2^{\lambda-m}}\\{i\:\text{odd}}}}U^{odd}_{i},\bigcup_{\substack{{1 \leq i \leq 2^{\lambda-m}}\\{i\:\text{even}}}}U^{odd}_{i},\bigcup_{\substack{{1 \leq i \leq 2^{\lambda-m}}\\{i\:\text{odd}}}}U^{even}_{i},\bigcup_{\substack{{1 \leq i \leq 2^{\lambda-m}}\\{i\:\text{even}}}}U^{even}_{i}},\\
&{\bigcup_{\substack{{1 \leq i \leq 2^{\lambda-m}}\\{i\:\text{odd}}}}V^{odd}_{i},\bigcup_{\substack{{1 \leq i \leq 2^{\lambda-m}}\\{i\:\text{even}}}}V^{odd}_{i},\bigcup_{\substack{{1 \leq i \leq 2^{\lambda-m}}\\{i\:\text{odd}}}}V^{even}_{i},\bigcup_{\substack{{1 \leq i \leq 2^{\lambda-m}}\\{i\:\text{even}}}}}V^{even}_{i}.
\end{align*}
}
\begin{proof}
From Lemma \ref{lemma1}, for $i\leq M$ to be adjacent to $j \leq M,$ $i$ and $j$ should differ by an odd number or an odd multiple of $2^{m}.$ It can be verified that no two elements which belong to the set $\displaystyle{\bigcup_{\substack{{1 \leq i \leq 2^{\lambda-m}}\\{i\:\text{odd}}}}U^{odd}_{i}}$ differ by an odd number or an odd multiple of $2^m.$ Hence, the elements of the set $\displaystyle{\bigcup_{\substack{{1 \leq i \leq 2^{\lambda-m}}\\{i\:\text{odd}}}}U^{odd}_{i}}$ can be colored using the same color. Similarly, the vertices which belong to the other 7 sets given in the statement of the lemma can be colored using the same color.
\end{proof}
\end{lemma}

Lemma \ref{lemma8} and Lemma \ref{lemma9} below provide proper 4-colorings for the cases when the singular fade states have absolute values $\sin(\pi k/M)$ or $\frac{1}{\sin(\pi k/M)}.$
\begin{lemma}
\label{lemma8}
Consider a singular fade state of the form $s=\sin(\pi k/M)e^{j\pi/M}$ or $s=\frac{1}{\sin(\pi k/M)}e^{j\pi/M},$ where $k$ is odd. A proper 4-coloring for $\mathcal{G}^V_s$ can be obtained by assigning 4 different colors to the vertices which belong to the following 4 sets: $\{1,3,\dotso M/2-1\},\{2,4,\dotso,M/2\},\{M/2+1,M/2+3,\dotso,M-1\},\{M/2+2,M/2+4,\dotso,M\}.$
\begin{proof}
From Lemma \ref{lemma3}, it follows that a vertex $i$ can be adjacent to vertex $j$ only when $i$ and $j$ differ by an odd number or $M/2.$ Hence the vertices which belong to one of the four sets in the statement of the lemma are not adjacent and hence can be colored using the same color. 
\end{proof}
\end{lemma}
\begin{lemma}
\label{lemma9}
Consider a singular fade state of the form $s=\sin(\pi k/M)$ or $s=\frac{1}{\sin(\pi k/M)},$ where $k=2^{m}a_1,$ $a_1$ being an odd integer. 
A proper 4-coloring for $\mathcal{G}^V_s$ can be obtained by assigning 4 different colors to the vertices which belong to the following 4 sets: 

{\footnotesize
\begin{align*}
&{\bigcup_{\substack{{1 \leq i \leq 2^{\lambda-m-1}}\\{i\:\text{odd}}}}\hspace{-.35 cm}U_{i},
\bigcup_{\substack{{1 \leq i \leq 2^{\lambda-m-1}}\\{i\:\text{even}}}}\hspace{-.35 cm}U_{i},
\bigcup_{\substack{{2^{\lambda-m-1}+1 \leq i \leq 2^{\lambda-m}}\\{i\:\text{odd}}}}\hspace{-.35 cm}U_{i},
\bigcup_{\substack{{2^{\lambda-m-1}+1 \leq i \leq 2^{\lambda-m}}\\{i\:\text{even}}}}\hspace{-.35 cm}U_{i}}.
\end{align*}
}
\begin{proof}
From Lemma \ref{lemma3}, it follows that a vertex $i$ can be adjacent to vertex $j$ only when $i$ and $j$ differ by an odd multiple of $2^{m}$ or $M/2.$ Hence the vertices which belong to one of the four sets in the statement of the lemma are not adjacent and hence can be colored using the same color. 
\end{proof}
\end{lemma}

From the $k$-coloring obtained for the vital subgraph of a singularity removal graph, a partially filled Latin Square with $k$ symbols can be obtained using the procedure described in Section III-B.
\subsection{Completion of Partially filled Latin Squares using $M$ symbols}
Before showing that the partially filled Latin Squares obtained in the previous subsection are completable using $M$ symbols, we state some useful results.

\begin{theorem}[\cite{MH}]
\label{thm_Hall}
Given a partially filled Latin Square of order $M$ with $M$ symbols, in which $r$ rows are completely filled, the rest of the empty $M-r$ rows can be filled with $M$ symbols to obtain a completely filled Latin Square.  
\end{theorem}
\begin{figure}
\centering
\subfigure[]{
\begin{tabular}{|c|c|c|c|}
\hline 1&2&3&4\\
\hline 3&1&4&2\\
\hline &&&\\
\hline &&&\\
\hline
\end{tabular}
\label{Hall_PFLS}
}
\subfigure[]{
\begin{tabular}{|c|c|c|c|}
\hline 1&2&3&4\\
\hline 3&1&4&2\\
\hline 4&3&2&1\\
\hline 2&4&1&3\\
\hline
\end{tabular}
\label{Hall_complete}
}
\caption{A partially filled Latin Square of order 4 with 2 completely filled rows and its completion with 4 symbols.}
\label{Hall_illustration}
\end{figure}

\begin{example}
\label{example_hall}
For example, the Latin Square of order 4 shown in Fig. \ref{Hall_PFLS}, which has two completely filled rows using 4 symbols, can be completed to form a Latin Square with 4 symbols, as shown in Fig. \ref{Hall_complete}.
\begin{figure}
\centering
\subfigure[]{
\begin{tabular}{|c|c|c|c|}
\hline 1&2&&\\
\hline &1&&2\\
\hline 2&&1&\\
\hline &&2&1\\
\hline
\end{tabular}
\label{corollary1_example}
}
\subfigure[]{
\begin{tabular}{|c|c|c|c|}
\hline 1&2&4&3\\
\hline 4&1&3&2\\
\hline 2&3&1&4\\
\hline 3&4&2&1\\
\hline
\end{tabular}
\label{corollary1_example_complete}
}
\caption{An example illustrating Corollary 1.}
\label{Hall_illustration}
\end{figure}
\end{example}

From a Latin Square filled with $M$ symbols, interchanging the symbol and row (column) indices results in another Latin Square.  Hence, a partially filled Latin Square $L$ with $M$ symbols is completable with $M$ symbols if and only if the partially filled Latin Square $L'$ obtained by interchanging the symbol and row (column) indices of the filled cells in $L$ is completable using $M$ symbols. A consequence of the above fact and Theorem \ref{thm_Hall} is stated in the following corollary. 
\begin{corollary}
\label{corollary1}
A partially filled Latin Square of order $M,$ in which only $s < M$ symbols are used and each one of the $s$ symbols appears in $M$ cells, can be completed using $M$ symbols.
\end{corollary}
\begin{example}
Continuing with Example \ref{example_hall}, interchanging the symbol and row indices of the filled cells in the partially filled Latin Square in Fig. \ref{Hall_PFLS} results in the partially filled Latin Square in Fig. \ref{corollary1_example}. From Corollary \ref{corollary1}, it follows that the partially filled Latin Square in Fig. \ref{corollary1_example} is completable using 4 symbols. One possible completion is shown in Fig. \ref{corollary1_example_complete}, which can also be obtained by interchanging the symbol and row indices in the Latin Square given in Fig. \ref{Hall_complete}. 
\end{example}

Let $X=\{X_1,X_2,\dotso ,X_n\}$ be a collection of subsets of a set $Y.$ A System of Distinct Representatives (SDR) for $X$ is a set of distinct elements $x_1,x_2,\dotso,x_n$ in $Y$ such that $x_i \in X_i.$ The element $x_i$ is said to be the representative of $X_i.$

\begin{example}
For $Y=\{1,2,3,4\},$ $X_1=\{1,2\},X_2=\{3,4\},X_3=\{1,2,3\},$ the set of numbers $1\in X_1,$ $3\in X_2$ and $2\in X_3$ form an SDR for $X=\{X_1,X_2,X_3\}.$
\end{example}

The following result which was shown by P. Hall, is known as Hall's marriage theorem.

\begin{theorem}[\cite{Ha_Ma}]
\label{thm_marriage}
An SDR for $X$ exists if and only if $\displaystyle{\left\vert\bigcup_{i \in S}X_i\right\vert}\geq \vert S \vert,$ for all $S \subseteq \{1,2,\dotso,n\}.$ 
\end{theorem}

Consider a singular fade state $s=\frac{\sin(\pi k/M)}{\sin(\pi l/M)}$ for which (i) both $k$ and $l$ are odd or (ii) $k$ and $l$ are respectively of the form $2^{m}a_1$ and $2^{m}a_2,$ where $a_1$ and $a_2$ are odd. The partially filled Latin Squares obtained for these two cases using the colorings provided in Lemma \ref{lemma4} and Lemma \ref{lemma5} respectively contain 4 symbols with each one of the 4 symbols appearing in $M$ cells. Hence, from Corollary \ref{corollary1}, it follows that these Latin Squares are completable using $M$ symbols and this is stated in the following theorem.    

\begin{theorem}
\label{thm_completable_1}
Consider a singular fade state $s=\frac{\sin(\pi k/M)}{\sin(\pi l/M)}$ for which (i) both $k$ and $l$ are odd or (ii) $k$ and $l$ are of the form $2^{m}a_1$ and $2^{m}a_2,$ where $a_1$ and $a_2$ are odd. The Latin Squares obtained for these two cases using the colorings provided in Lemma \ref{lemma4} and Lemma \ref{lemma5} respectively are completable using $M$ symbols. 
\end{theorem}

An explicit way exists for filling the empty cells of the partially filled Latin Squares considered in Theorem \ref{thm_completable_1}. Note that all the partially filled Latin Squares obtained for those singular fade states with absolute value $\frac{\sin(\pi k/M)}{\sin(\pi l/M)}$ for which $k,l\neq M/2$ have 4 filled cells in every column. Let $i_1,i_2,i_3,i_4,\dotso i_{M-4}$ denote the column indices of the cells which are empty in the first row. Note that the empty cells in the partially filled Latin Squares are of the form $\left(1+b,\left(i_l-1+b\right)\text{mod}\: M+1\right),b\in\{0,1,\dotso M-1\},l\in\{1,2,\dotso,M-4\}.$ Also, note that the partially filled Latin Squares considered in Theorem \ref{thm_completable_1} contain only 4 symbols. Starting from the $(1,i_l)$-th cell and filling diagonally with the symbol $4+l,$ we can get a completely filled Latin Square. In other words, for $l\in\{1,2,\dotso,M-4\},$ fill all the cells which belong to the set $\{\left(1+b,\left(i_l-1+b\right)\text{mod}\: M+1\right),b\in\{0,1,\dotso M-1\}\}$ with the symbol $4+l$ to get a completely filled Latin Square. This is illustrated in the following examples.
\begin{example}
For 8-PSK signal set consider the singular fade state $s=\frac{\sin(\pi/8)}{\sin(3\pi/8)}.$ The constrained partial Latin Square for this case is as shown in Fig. \ref{example_thm3_cpls}. Using the coloring given in Lemma \ref{example_thm3_cpls}, the partially filled Latin Square given in Fig. \ref{example_thm3_pfls} is obtained. In the partially filled Latin Square in Fig. \ref{example_thm3_pfls}, the empty cells in the first row have column indices 1, 2, 5 and 8. Starting from the cells $(1,1),(1,2),(1,5),(1,8)$ and filling along the diagonal with the symbols 5, 6, 7 and 8 respectively, the completely filled Latin Square shown in Fig. \ref{example_thm3_ls} is obtained.  
\begin{figure}
\centering
\subfigure[]{
\begin{tabular}{|c|c|c|c|c|c|c|c|}
\hline  &     &    10 &    2  &          &     1  &    9  &    \\
\hline  &     &       &    11 &    3     &        &    2  &  10\\
\hline 11     &     &   &  &    12 &    4     & &   3\\
\hline  4   & 12     &     &    & &    13  &   5     &\\
\hline   &     5  &  13     &     &     &   & 14 &    6\\
\hline  7     &    & 6&    14     &     &     &  &  15\\
\hline 16   &  8     & &    7  &  15     &     &     &\\
\hline   &     9 &    1     &   &  8&    16     &     &\\
\hline
\end{tabular}
\label{example_thm3_cpls}
}
\subfigure[]{
\begin{tabular}{|c|c|c|c|c|c|c|c|}
\hline  &     &    4 &    2  &          &     1  &    3  &    \\
\hline  &     &       &    3 &    1     &        &    2  &  4\\
\hline 3     &     &   &  &    4 &    2     & &   1\\
\hline  2   & 4     &     &    & &    3  &   1     &\\
\hline   &     1  &  3     &     &     &   & 4 &    2\\
\hline  1     &    & 2&    4     &     &     &  &  3\\
\hline 4   &  2     & &    1 &  3     &     &     &\\
\hline   &     3 &    1     &   &  2&    4     &     &\\
\hline
\end{tabular}
\label{example_thm3_pfls}
}
\subfigure[]{
\begin{tabular}{|c|c|c|c|c|c|c|c|}
\hline  5& 6    &    4 &    2  & 7         &     1  &    3  &  8  \\
\hline 8 &5     &6       &    3 &    1     &7        &    2  &  4\\
\hline 3     & 8    &5   &6  &    4 &    2     &7 &   1\\
\hline  2   & 4     & 8    &5    &6 &    3  &   1     &7\\
\hline  7 &     1  &  3     &8     &5     & 6  & 4 &    2\\
\hline  1     &7    & 2&    4     &8     &5     &6  &  3\\
\hline 4   &  2     &7 &    1 &  3     &8     &5     &6\\
\hline 6  &     3 &    1     &7   &  2&    4     &8     &5\\
\hline
\end{tabular}
\label{example_thm3_ls}
}
\caption{Constrained partial Latin Square, partially filled Latin Square and Latin Square obtained for the singular fade state $\frac{\sin(\pi/8)}{\sin(3\pi/8)}.$}
\label{example_thm3}
\end{figure}
\end{example}
\begin{example}
Consider the singular fade state $\frac{\sin(2\pi/16)}{\sin(6\pi/16)}$ for 16-PSK signal set. For this case, The constrained partial Latin Square and the partially filled Latin Square obtained using the 4-coloring provided in Lemma \ref{lemma5} are shown respectively in Fig. \ref{example2_thm3_cpls} and Fig. \ref{example2_thm3_pfls}. The partially filled Latin Square in Fig. \ref{example2_thm3_pfls} can be completed using the diagonal filling approach described earlier to get the Latin Square given in Fig. \ref{example2_thm3_ls}. 
\begin{figure*}
\centering
\subfigure[]{\footnotesize
\begin{tabular}{|c|c|c|c|c|c|c|c|c|c|c|c|c|c|c|c|}
\hline       &     &        &     &   19  &    &     3&      &      &      &     1&      &    17&      &      &      \\
\hline      &     &      &      &      &    20&      &     4 &     &      &      &     2 &     &    18&      &    \\
\hline      &      &      &      &      &      &    21&      &     5&      &      &      &     3    &  &    19&      \\
\hline      &      &      &      &      &      &      &    22 &     &     6  &    &      &      &     4 &     &    20\\
\hline    21 &     &      &      &      &      &      &      &    23 &     &     7&      &      &      &     5&      \\
\hline      &    22  &    &      &      &      &      &      &      &    24&      &     8&      &      &      &     6\\
\hline     7  &    &    23&      &      &      &      &      &      &      &    25 &     &     9&      &      &      \\
 \hline     &     8 &     &    24   &   &      &      &      &      &      &      &    26&      &    10   &   &      \\
\hline      &      &     9   &   &    25&      &      &      &      &      &      &      &    27&      &    11&      \\
\hline      &      &      &    10 &     &    26&      &      &      &      &      &      &      &    28&      &    12\\
\hline    13   &   &      &      &    11&      &    27  &    &      &      &      &      &      &      &    29 &     \\
\hline      &    14  &    &      &      &    12 &     &    28  &    &      &      &      &      &      &      &    30\\
\hline    31 &     &    15 &     &      &      &    13&      &    29&      &      &      &      &      &      &      \\
\hline      &    32 &     &    16&      &      &      &    14 &     &    30&      &      &      &      &      &      \\
 \hline     &      &    17 &     &     1&      &      &      &    15&      &    31&      &      &      &      &      \\
\hline      &      &      &    18 &     &     2 &     &      &      &    16&      &    32&      &      &      &      \\
\hline
\end{tabular}
\label{example2_thm3_cpls}
}
\subfigure[]{\footnotesize
\begin{tabular}{|c|c|c|c|c|c|c|c|c|c|c|c|c|c|c|c|}
\hline       &     &        &     &   4  &    &     2&      &      &      &     1&      &    3&      &      &      \\
\hline      &     &      &      &      &    4&      &     2 &     &      &      &     1&     &    3&      &    \\
\hline      &      &      &      &      &      &    3&      &     1&      &      &      &     2    &  &    4&      \\
\hline      &      &      &      &      &      &      &    3 &     &     1  &    &      &      &     2 &     &    4\\
\hline    3 &     &      &      &      &      &      &      &    4 &     &     2&      &      &      &     1&      \\
\hline      &    3  &    &      &      &      &      &      &      &    4&      &     2&      &      &      &     1\\
\hline     2  &    &    4&      &      &      &      &      &      &      &    3 &     &     1&      &      &      \\
 \hline     &     2 &     &    4  &   &      &      &      &      &      &      &    3&      &    1   &   &      \\
\hline      &      &     1   &   &    3&      &      &      &      &      &      &      &    4&      &    2&      \\
\hline      &      &      &    1 &     &    3&      &      &      &      &      &      &      &    4&      &    2\\
\hline    1   &   &      &      &    2&      &    4  &    &      &      &      &      &      &      &    3 &     \\
\hline      &    1  &    &      &      &    2 &     &    4  &    &      &      &      &      &      &      &    3\\
\hline    4 &     &    2 &     &      &      &    1&      &    3&      &      &      &      &      &      &      \\
\hline      &    4 &     &   2&      &      &      &    1 &     &    3&      &      &      &      &      &      \\
 \hline     &      &    3 &     &     1&      &      &      &    2&      &    4&      &      &      &      &      \\
\hline      &      &      &    3 &     &     1 &     &      &      &    2&      &    4&      &      &      &      \\
\hline
\end{tabular}
\label{example2_thm3_pfls}
}
\subfigure[]{\footnotesize
\begin{tabular}{|c|c|c|c|c|c|c|c|c|c|c|c|c|c|c|c|}
\hline     5 &    6&     7  &   8 &   4  &   9&     2&     10&     11&     12&     1&     13&    3&     14&     15&     16\\
\hline     16&    5&     6&     7&     8&    4&     9&     2 &    10&     11&     12&     1&    13&    3&     14&   15\\
\hline     15&     16&     5&     6&     7&     8&    3&     9&     1&     10&     11&     12&     2    & 13&    4&     14\\
\hline     14&     15&     16&     5&     6&     7&     8&    3 &    9&     1  &   10&     11&     12&     2 &    13&    4\\
\hline    3 &    14&     15&     16&     5&     6&     7&     8&    4 &    9&     2&     10&     11&     12&     1&     13\\
\hline     13&    3  &   14&     15&     16&    5&     6&     7&     8&    4&     9&     2&     10&     11&     12&     1\\
\hline     2  &   13&    4&     14&     15&     16&     5&     6&     7&     8&    3 &    9&     1&     10&     11&     12\\
 \hline    12&     2 &    13&    4  &  14&     15&     16&     5&     6&     7&     8&    3&     9&    1   &  10&     11\\
\hline     11&     12&     1   &  13&    3&     14&     15&     16&     5&     6&     7&     8&    4&     9&    2&     10\\
\hline     10&     11&     12&    1 &    13&    3&     14&     15&     16&     5&     6&     7&     8&    4&     9&    2\\
\hline    1   &  10&     11&     12&    2&     13&    4  &   14&     15&     16&     5&     6&     7&     8&    3 &    9\\
\hline     9&    1  &   10&     11&     12&    2 &    13&    4  &   14&     15&     16&     5&     6&     7&     8&    3\\
\hline    4 &    9&    2 &    10&     11&     12&    1&     13&    3&     14&     15&     16&     5&     6&     7&     8\\
\hline     8&    4 &    9&   2&     10&     11&     12&    1 &    13&    3&     14&     15&     16&     5&     6&     7\\
 \hline    7&     8&    3 &    9&     1&     10&     11&     12&    2&     13&    4&     14&     15&     16&     5&     6\\
\hline     6&     7&     8&    3 &    9&     1 &    10&     11&     12&    2&     13&    4&     14&     15&    16&     5\\
\hline
\end{tabular}
\label{example2_thm3_ls}
}
\caption{Constrained partial Latin Square, partially filled Latin Square and Latin Square obtained for the singular fade state $\frac{\sin(2\pi/16)}{\sin(3\pi/16)}.$}
\end{figure*}
\end{example}

For a singular fade state whose absolute value is $\sin(\pi k/M)$ or $\frac{1}{\sin(\pi k/M)},$ the 4-colorings provided in Lemma \ref{lemma8} and Lemma \ref{lemma9} result in partially filled Latin Squares which are filled in with 4 symbols, with each symbol appearing in $M/2$ cells and two cells filled in every row and every column. In Theorem \ref{thm4} below, the completability of such Latin Squares is proved. 
\begin{theorem}
\label{thm4}
Consider a singular fade state whose absolute value is $\sin(\pi k/M)$ or $\frac{1}{\sin(\pi k/M)}.$ For this case, the partially filled Latin Squares obtained using the 4-colorings provided in Lemma \ref{lemma8} and Lemma \ref{lemma9} are completable using $M$ symbols.
\begin{proof}
See Appendix \ref{appendix1}.
\end{proof}
\end{theorem}

The ideas used in the proof of Theorem \ref{thm4} are illustrated in the following example.
\begin{example}
Consider the singular fade state $s=\sin(2\pi/8)$ for 8-PSK signal set. The constrained partial Latin Square for this case is shown in Fig. \ref{example_thm4_cpls}. Using the 4-coloring given in Lemma \ref{lemma9}, the partially filled Latin Square $L$ given in Fig. \ref{example_thm4_pfls} is obtained. 
Note that in $L,$ two cells are filled in every row and every column.
In the partially filled Latin Square $L,$ in each row, one more cell is filled in with the symbols 1 to 4, to obtain the partially filled Latin Square $L_1$ given in Fig. \ref{example_thm4_pfls_2}. The partially filled Latin Square $L'_1$ given in Fig. \ref{example_thm4_pfls_3} is obtained by interchanging the symbol and row indices of the filled cells in the partially filled Latin Square $L_1.$ Consider the partially filled Latin Rectangle $L'_R$ obtained by taking only the first 4 rows of $L'_1.$
Let $X_i, i \in \{1,2,\dotso,8\}$ denote the set of cells in which symbol $i$ can appear in $L'_R.$ From Fig. \ref{example_thm4_pfls_3}, the sets $X_i, i \in \{1,2,\dotso,8\}$ can be obtained as follows: $X_1=\{(4,7)\},X_2=\{(4,6)\},X_3=\{(1,1)\},X_4=\{(1,8)\},X_5=\{(2,3)\},X_6=\{(2,2)\},X_7=\{(3,5)\},$ and $X_8=\{(3,4)\}.$ There exists an SDR for $X=\{X_1,X_2,\dotso,X_8\},$ with the only element of $X_i$ being its representative. The partially filled Latin Square in Fig. \ref{example_thm4_pfls_4} is obtained from $L'_R$ by filling the representative of $X_i$ with the symbol $i.$ 
From Theorem \ref{thm_Hall}, it follows that the partially filled Latin Square in Fig. \ref{example_thm4_pfls_4} is completable using $8$ symbols. Fig. \ref{example_thm4_ls1} shows one such completion. By interchanging the symbol and row indices in the Latin Square in Fig. \ref{example_thm4_ls1}, we get the Latin Square given in Fig. \ref{example_thm4_ls2} which removes the singular fade state $\sin(2\pi/8)$ for 8-PSK signal set.

\begin{figure}
\centering
\subfigure[]{\tiny
\begin{tabular}{|c|c|c|c|c|c|c|c|}
\hline  &     &     &      3&          &      1&     &    \\
\hline  &     &       &     &         4&        &    2  &  \\
\hline      &     &   &  &    &    5     & &   3\\
\hline  4   &      &     &    & &     &   6    &\\
\hline   &     5  &       &     &     &   & &    7\\
\hline  8     &    & 6&         &     &     &  &  \\
\hline    &  1     & &    7  &       &     &     &\\
\hline   &      &    2     &   &  8&         &     &\\
\hline
\end{tabular}
\label{example_thm4_cpls}
}
\subfigure[]{\tiny
\begin{tabular}{|c|c|c|c|c|c|c|c|}
\hline • & • & • & 2 & • & 1 & • & •  \\ 
\hline • & • & • & • & 2 & • & 1 & • \\ 
\hline • & • & • & • &  & 3 &  • & 2 \\ 
\hline 2 & • & • & • &  & • &  3 & • \\ 
\hline • & 3 & • & • & • & • & • & 4 \\ 
\hline 4 & • & 3 & • & • & • & • & • \\ 
\hline • & 1 & • & 4 & • & • & • & • \\ 
\hline • & • & 1 & • & 4 & • & • & • \\ 
\hline 
\end{tabular} 
\label{example_thm4_pfls}
}
\subfigure[]{\tiny
\begin{tabular}{|c|c|c|c|c|c|c|c|}
\hline • & • & • & 2 & • & 1 & • & 3 \\ 
\hline 3 & • & • & • & 2 & • & 1 & • \\ 
\hline • & 4 & • & • & • & 3 & • & 2 \\ 
\hline 2 & • & 4 & • & • & • & 3 & • \\ 
\hline • & 3 & • & 1 & • & • & • & 4 \\ 
\hline 4 & • & 3 & • & 1 & • & • & • \\ 
\hline • & 1 & • & 4 & • & 2 & • & • \\ 
\hline • & • & 1 & • & 4 & • & 2 & • \\ 
\hline 
\end{tabular} 
\label{example_thm4_pfls_2}
}
\subfigure[]{\tiny
\begin{tabular}{|c|c|c|c|c|c|c|c|}
\hline
• & 7  & 8  & 5  & 6  & 1  & 2  & • \\
\hline 4  & • & • & 1  & 2  & 7  & 8  & 3  \\
\hline 2  & 5  & 6  & • & • & 3  & 4  & 1  \\
\hline 6  & 3  & 4  & 7  & 8  & • & • & 5  \\
\hline • & • & • & • & • & • & • & • \\
\hline • & • & • & • & • & • & • & • \\
\hline • & • & • & • & • & • & • & • \\
\hline • & • & • & • & • & • & • & • \\
\hline

\end{tabular} 
\label{example_thm4_pfls_3}
}
\subfigure[]{\tiny
\begin{tabular}{|c|c|c|c|c|c|c|c|}
\hline
3 & 7  & 8  & 5  & 6  & 1  & 2  & 4 \\
\hline 4  & 6 & 5 & 1  & 2  & 7  & 8  & 3  \\
\hline 2  & 5  & 6  & 8 & 7 & 3  & 4  & 1  \\
\hline 6  & 3  & 4  & 7  & 8  & 2 & 1 & 5  \\
\hline • & • & • & • & • & • & • & • \\
\hline • & • & • & • & • & • & • & • \\
\hline • & • & • & • & • & • & • & • \\
\hline • & • & • & • & • & • & • & • \\
\hline

\end{tabular} 
\label{example_thm4_pfls_4}
}
\subfigure[]{\tiny
\begin{tabular}{|c|c|c|c|c|c|c|c|}
\hline
3  & 7  & 8  & 5  & 6  & 1  & 2  & 4  \\
\hline
4  & 6  & 5  & 1  & 2  & 7  & 8  & 3  \\
\hline
2  & 5  & 6  & 8  & 7  & 3  & 4  & 1  \\
\hline
6  & 3  & 4  & 7  & 8  & 2  & 1  & 5  \\
\hline
1  & 2  & 3  & 4  & 5  & 6  & 7  & 8  \\
\hline
8  & 1  & 2  & 3  & 4  & 5  & 6  & 7  \\
\hline
7  & 8  & 1  & 2  & 3  & 4  & 5  & 6  \\
\hline
5  & 4  & 7  & 6  & 1  & 8  & 3  & 2  \\
\hline

\end{tabular} 
\label{example_thm4_ls1}
}
\subfigure[]{\tiny
\begin{tabular}{|c|c|c|c|c|c|c|c|}
\hline 5 & 6 & 7 & 2 & 8 & 1 & 4 & 3 \\ 
\hline 3 & 5 & 6 & 7 & 2 & 4 & 1 & 8 \\ 
\hline 1 & 4 & 5 & 6 & 7 & 3 & 8 & 2 \\ 
\hline 2 & 8 & 4 & 5 & 6 & 7 & 3 & 1 \\ 
\hline 8 & 3 & 2 & 1 & 5 & 6 & 7 & 4 \\ 
\hline 4 & 2 & 3 & 8 & 1 & 5 & 6 & 7 \\ 
\hline 7 & 1 & 8 & 4 & 3 & 2 & 5 & 6 \\ 
\hline 6 & 7 & 1 & 3 & 4 & 8 & 2 & 5 \\ 
\hline 
\end{tabular}
\label{example_thm4_ls2}
}
\caption{Example illustrating the ideas used in the proof of Theorem \ref{thm4}.}
\label{example_thm3}
\end{figure}
\end{example}

For those singular fade states of the form $\frac{\sin(\pi k/M)}{\sin(\pi l/M)},$ $k,l\neq M/2,$ excluding the ones considered in Theorem \ref{thm_completable_1}, the partially filled Latin Squares obtained using the 8-colorings provided in Lemma \ref{lemma6} and Lemma \ref{lemma7} contain 8 symbols, with each symbol appearing in $M/2$ cells. Each row (column) of these partially filled Latin Squares contains 4 filled cells. In the following theorem, it is shown that these partially filled Latin Squares are completable using $M$ symbols.

\begin{theorem}
\label{thm_completable_3}
Consider a singular fade state $s=\frac{\sin(\pi k/M)}{\sin(\pi l/M)}$ for which (ii) $k$ and $l$ are of the form $2^{m_1}a_1$ and $2^{m_2}a_2,$ where $a_1$ and $a_2$ are odd and $m_1 < m_2$ or (ii) only one among $k$ and $l$ is odd. The Latin Squares obtained for these two cases using the 8-colorings provided in Lemma \ref{lemma6} and Lemma \ref{lemma7} respectively are completable using $M$ symbols. 
\begin{proof}
See Appendix \ref{appendix2}.
\end{proof}
\end{theorem}

The ideas used in the proof of Theorem \ref{thm_completable_3} are illustrated in the following example.
\begin{example}
\begin{figure*}
\centering
\subfigure[]{\tiny
\begin{tabular}{|c|c|c|c|c|c|c|c|c|c|c|c|c|c|c|c|}
\hline
•   & 18   & •   & •   & •   & •   & •   & 2   & 1   & •   & •   & •   & •   & •   & 17   & •   \\
\hline
•   & •   & 19   & •   & •   & •   & •   & •   & 3   & 2   & •   & •   & •   & •   & •   & 18   \\
\hline
19   & •   & •   & 20   & •   & •   & •   & •   & •   & 4   & 3   & •   & •   & •   & •   & •   \\
\hline
•   & 20   & •   & •   & 21   & •   & •   & •   & •   & •   & 5   & 4   & •   & •   & •   & •   \\
\hline
•   & •   & 21   & •   & •   & 22   & •   & •   & •   & •   & •   & 6   & 5   & •   & •   & •   \\
\hline
•   & •   & •   & 22   & •   & •   & 23   & •   & •   & •   & •   & •   & 7   & 6   & •   & •   \\
\hline
•   & •   & •   & •   & 23   & •   & •   & 24   & •   & •   & •   & •   & •   & 8   & 7   & •   \\
\hline
•   & •   & •   & •   & •   & 24   & •   & •   & 25   & •   & •   & •   & •   & •   & 9   & 8   \\
\hline
9   & •   & •   & •   & •   & •   & 25   & •   & •   & 26   & •   & •   & •   & •   & •   & 10   \\
\hline
11   & 10   & •   & •   & •   & •   & •   & 26   & •   & •   & 27   & •   & •   & •   & •   & •   \\
\hline
•   & 12   & 11   & •   & •   & •   & •   & •   & 27   & •   & •   & 28   & •   & •   & •   & •   \\
\hline
•   & •   & 13   & 12   & •   & •   & •   & •   & •   & 28   & •   & •   & 29   & •   & •   & •   \\
\hline
•   & •   & •   & 14   & 13   & •   & •   & •   & •   & •   & 29   & •   & •   & 30   & •   & •   \\
\hline
•   & •   & •   & •   & 15   & 14   & •   & •   & •   & •   & •   & 30   & •   & •   & 31   & •   \\
\hline
•   & •   & •   & •   & •   & 16   & 15   & •   & •   & •   & •   & •   & 31   & •   & •   & 32   \\
\hline
17   & •   & •   & •   & •   & •   & 1   & 16   & •   & •   & •   & •   & •   & 32   & •   & • \\
\hline

\hline
\end{tabular}
\label{example_thm6_cpls}
}
\subfigure[]{\tiny
\begin{tabular}{|c|c|c|c|c|c|c|c|c|c|c|c|c|c|c|c|}
\hline • & 6  & • & • & • & • & • & 2  & 1  & • & • & • & • & • & 5  & • \\
\hline • & • & 7  & • & • & • & • & • & 3  & 2  & • & • & • & • & • & 6  \\
\hline 7  & • & • & 8  & • & • & • & • & • & 4  & 3  & • & • & • & • & • \\
\hline • & 8  & • & • & 5  & • & • & • & • & • & 1  & 4  & • & • & • & • \\
\hline • & • & 5  & • & • & 6  & • & • & • & • & • & 2  & 1  & • & • & • \\
\hline • & • & • & 6  & • & • & 7  & • & • & • & • & • & 3  & 2  & • & • \\
\hline • & • & • & • & 7  & • & • & 8  & • & • & • & • & • & 4  & 3  & • \\
\hline • & • & • & • & • & 8  & • & • & 5  & • & • & • & • & • & 1  & 4  \\
\hline 1  & • & • & • & • & • & 5  & • & • & 6  & • & • & • & • & • & 2  \\
\hline 3  & 2  & • & • & • & • & • & 6  & • & • & 7  & • & • & • & • & • \\
\hline • & 4  & 3  & • & • & • & • & • & 7  & • & • & 8  & • & • & • & • \\
\hline • & • & 1  & 4  & • & • & • & • & • & 8  & • & • & 5  & • & • & • \\
\hline • & • & • & 2  & 1  & • & • & • & • & • & 5  & • & • & 6  & • & • \\
\hline • & • & • & • & 3  & 2  & • & • & • & • & • & 6  & • & • & 7  & • \\
\hline • & • & • & • & • & 4  & 3  & • & • & • & • & • & 7  & • & • & 8  \\
\hline 5  & • & • & • & • & • & 1  & 4  & • & • & • & • & • & 8  & • & • \\
\hline
\end{tabular} 
\label{example_thm6_pfls}
}
\subfigure[]{\tiny
\begin{tabular}{|c|c|c|c|c|c|c|c|c|c|c|c|c|c|c|c|}
\hline
4   & 6   &  • &  • &  • &  • & 8   & 2   & 1   &  • &  • &  • &  • &  • & 5   &  • \\
\hline  • & 1   & 7   &  • &  • &  • &  • & 5   & 3   & 2   &  • &  • &  • &  • &  • & 6   \\
\hline 7   &  • & 2   & 8   &  • &  • &  • &  • & 6   & 4   & 3   &  • &  • &  • &  • &  • \\
\hline  • & 8   &  • & 3   & 5   &  • &  • &  • &  • & 7   & 1   & 4   &  • &  • &  • &  • \\
\hline  • &  • & 5   &  • & 4   & 6   &  • &  • &  • &  • & 8   & 2   & 1   &  • &  • &  • \\
\hline  • &  • &  • & 6   &  • & 1   & 7   &  • &  • &  • &  • & 5   & 3   & 2   &  • &  • \\
\hline  • &  • &  • &  • & 7   &  • & 2   & 8   &  • &  • &  • &  • & 6   & 4   & 3   &  • \\
\hline  • &  • &  • &  • &  • & 8   &  • & 3   & 5   &  • &  • &  • &  • & 7   & 1   & 4   \\
\hline 1   &  • &  • &  • &  • &  • & 5   &  • & 4   & 6   &  • &  • &  • &  • & 8   & 2   \\
\hline 3   & 2   &  • &  • &  • &  • &  • & 6   &  • & 1   & 7   &  • &  • &  • &  • & 5   \\
\hline 6   & 4   & 3   &  • &  • &  • &  • &  • & 7   &  • & 2   & 8   &  • &  • &  • &  • \\
\hline  • & 7   & 1   & 4   &  • &  • &  • &  • &  • & 8   &  • & 3   & 5   &  • &  • &  • \\
\hline  • &  • & 8   & 2   & 1   &  • &  • &  • &  • &  • & 5   &  • & 4   & 6   &  • &  • \\
\hline  • &  • &  • & 5   & 3   & 2   &  • &  • &  • &  • &  • & 6   &  • & 1   & 7   &  • \\
\hline  • &  • &  • &  • & 6   & 4   & 3   &  • &  • &  • &  • &  • & 7   &  • & 2   & 8   \\
\hline 5   &  • &  • &  • &  • & 7   & 1   & 4   &  • &  • &  • &  • &  • & 8   &  • & 3   \\
\hline
\end{tabular} 
\label{example_thm6_pfls_2}
}
\subfigure[]{\tiny
\begin{tabular}{|c|c|c|c|c|c|c|c|c|c|c|c|c|c|c|c|}
\hline
9   & 2   & 12   &   • & 13   & 6   & 16   &   • & 1   & 10  • & 4   &   • & 5   & 14   & 8   &   • \\
\hline
  • & 1  • & 3   & 13   &   • & 14   & 7   & 1   &   • & 2   & 11   & 5   &   • & 6   & 15   & 9   \\
\hline
10  • &   • & 11   & 4   & 14   &   • & 15   & 8   & 2   &   • & 3   & 12   & 6   &   • & 7   & 16   \\
\hline
1   & 11   &   • & 12   & 5   & 15   &   • & 16   & 9   & 3   &   • & 4   & 13   & 7   &   • & 8   \\
\hline
16   &   • & 5   & 14   & 4   &   • & 9   & 2   & 8   &   • & 13   & 6   & 12   &   • & 1   & 10  • \\
\hline
11   & 1   &   • & 6   & 15   & 5   &   • & 10  • & 3   & 9   &   • & 14   & 7   & 13   &   • & 2   \\
\hline
3   & 12   & 2   &   • & 7   & 16   & 6   &   • & 11   & 4   & 10  • &   • & 15   & 8   & 14   &   • \\
\hline
  • & 4   & 13   & 3   &   • & 8   & 1   & 7   &   • & 12   & 5   & 11   &   • & 16   & 9   & 15   \\
\hline
  • &   • &   • &   • &   • &   • &   • &   • &   • &   • &   • &   • &   • &   • &   • &   • \\
\hline
  • &   • &   • &   • &   • &   • &   • &   • &   • &   • &   • &   • &   • &   • &   • &   • \\
\hline
  • &   • &   • &   • &   • &   • &   • &   • &   • &   • &   • &   • &   • &   • &   • &   • \\
\hline
  • &   • &   • &   • &   • &   • &   • &   • &   • &   • &   • &   • &   • &   • &   • &   • \\
\hline
  • &   • &   • &   • &   • &   • &   • &   • &   • &   • &   • &   • &   • &   • &   • &   • \\
\hline
  • &   • &   • &   • &   • &   • &   • &   • &   • &   • &   • &   • &   • &   • &   • &   • \\
\hline
  • &   • &   • &   • &   • &   • &   • &   • &   • &   • &   • &   • &   • &   • &   • &   • \\
\hline
  • &   • &   • &   • &   • &   • &   • &   • &   • &   • &   • &   • &   • &   • &   • &   • \\
\hline

\end{tabular} 
\label{example_thm6_pfls_3}
}
\subfigure[]{\tiny
\begin{tabular}{|c|c|c|c|c|c|c|c|c|c|c|c|c|c|c|c|}
\hline
9   & 2   & 12   &   7 & 13   & 6   & 16   &   3 & 1   & 10   & 4   &   15 & 5   & 14   & 8   &   11 \\
\hline
  4 & 10 & 3   & 13   &   8 & 14   & 7   & 1   &   12 & 2   & 11   & 5   &   16 & 6   & 15   & 9   \\
\hline
10  &   5 & 11   & 4   & 14   &   1 & 15   & 8   & 2   &   13 & 3   & 12   & 6   &   9 & 7   & 16   \\
\hline
1   & 11   &   6 & 12   & 5   & 15   &   2 & 16   & 9   & 3   &   14& 4   & 13   & 7   &   10 & 8   \\
\hline
16   &   3 & 5   & 14   & 4   &   7 & 9   & 2   & 8   &   11& 13   & 6   & 12   &   15 & 1   & 10  \\
\hline
11   & 1   &   4 & 6   & 15   & 5   &   8 & 10   & 3   & 9   &   12 & 14   & 7   & 13   &   16 & 2   \\
\hline
3   & 12   & 2   &   1 & 7   & 16   & 6   &   5 & 11   & 4   & 10  • &   9 & 15   & 8   & 14   &   13\\
\hline
  2 & 4   & 13   & 3   &   6 & 8   & 1   & 7   &   10 & 12   & 5   & 11   &   14 & 16   & 9   & 15   \\
\hline
  • &   • &   • &   • &   • &   • &   • &   • &   • &   • &   • &   • &   • &   • &   • &   • \\
\hline
  • &   • &   • &   • &   • &   • &   • &   • &   • &   • &   • &   • &   • &   • &   • &   • \\
\hline
  • &   • &   • &   • &   • &   • &   • &   • &   • &   • &   • &   • &   • &   • &   • &   • \\
\hline
  • &   • &   • &   • &   • &   • &   • &   • &   • &   • &   • &   • &   • &   • &   • &   • \\
\hline
  • &   • &   • &   • &   • &   • &   • &   • &   • &   • &   • &   • &   • &   • &   • &   • \\
\hline
  • &   • &   • &   • &   • &   • &   • &   • &   • &   • &   • &   • &   • &   • &   • &   • \\
\hline
  • &   • &   • &   • &   • &   • &   • &   • &   • &   • &   • &   • &   • &   • &   • &   • \\
\hline
  • &   • &   • &   • &   • &   • &   • &   • &   • &   • &   • &   • &   • &   • &   • &   • \\
\hline
\end{tabular} 
\label{example_thm6_pfls_4}
}
\subfigure[]{\tiny
\begin{tabular}{|c|c|c|c|c|c|c|c|c|c|c|c|c|c|c|c|}
\hline
9    & 2    & 12    & 7    & 13    & 6    & 16    & 3    & 1    & 10    & 4    & 15    & 5    & 14    & 8    & 11    \\
\hline 4    & 10    & 3    & 13    & 8    & 14    & 7    & 1    & 12    & 2    & 11    & 5    & 16    & 6    & 15    & 9    \\
\hline 10    & 5    & 11    & 4    & 14    & 1    & 15    & 8    & 2    & 13    & 3    & 12    & 6    & 9    & 7    & 16    \\
\hline 1    & 11    & 6    & 12    & 5    & 15    & 2    & 16    & 9    & 3    & 14    & 4    & 13    & 7    & 10    & 8    \\
\hline 16    & 3    & 5    & 14    & 4    & 7    & 9    & 2    & 8    & 11    & 13    & 6    & 12    & 15    & 1    & 10    \\
\hline 11    & 1    & 4    & 6    & 15    & 5    & 8    & 10    & 3    & 9    & 12    & 14    & 7    & 13    & 16    & 2    \\
\hline 3    & 12    & 2    & 1    & 7    & 16    & 6    & 5    & 11    & 4    & 10    & 9    & 15    & 8    & 14    & 13    \\
\hline 2    & 4    & 13    & 3    & 6    & 8    & 1    & 7    & 10    & 12    & 5    & 11    & 14    & 16    & 9    & 15    \\
\hline 5    & 6    & 1    & 2    & 3    & 4    & 10    & 9    & 13    & 7    & 15    & 16    & 8    & 11    & 12    & 14    \\
\hline 6    & 7    & 8    & 5    & 1    & 2    & 3    & 13    & 14    & 15    & 16    & 10    & 9    & 4    & 11    & 12    \\
\hline 7    & 8    & 9    & 10    & 2    & 3    & 4    & 14    & 15    & 16    & 1    & 13    & 11    & 12    & 5    & 6    \\
\hline 8    & 9    & 7    & 11    & 10    & 12    & 5    & 15    & 16    & 14    & 6    & 1    & 2    & 3    & 13    & 4    \\
\hline 12    & 13    & 10    & 15    & 16    & 9    & 14    & 11    & 4    & 1    & 7    & 8    & 3    & 2    & 6    & 5    \\
\hline 13    & 14    & 15    & 16    & 9    & 10    & 11    & 12    & 5    & 6    & 8    & 2    & 4    & 1    & 3    & 7    \\
\hline 14    & 15    & 16    & 8    & 11    & 13    & 12    & 4    & 6    & 5    & 9    & 7    & 1    & 10    & 2    & 3    \\
\hline 15    & 16    & 14    & 9    & 12    & 11    & 13    & 6    & 7    & 8    & 2    & 3    & 10    & 5    & 4    & 1    \\
\hline

\end{tabular} 
\label{example_thm6_ls1}
}
\subfigure[]{\tiny
\begin{tabular}{|c|c|c|c|c|c|c|c|c|c|c|c|c|c|c|c|}
\hline
4    & 6    & 9    & 7    & 10    & 3    & 8    & 2    & 1    & 13    & 11    & 12    & 15    & 14    & 5    & 16    \\
\hline 8    & 1    & 7    & 9    & 11    & 10    & 4    & 5    & 3    & 2    & 16    & 14    & 12    & 13    & 15    & 6    \\
\hline 7    & 5    & 2    & 8    & 9    & 11    & 10    & 1    & 6    & 4    & 3    & 16    & 13    & 12    & 14    & 15    \\
\hline 2    & 8    & 6    & 3    & 5    & 9    & 11    & 15    & 13    & 7    & 1    & 4    & 14    & 10    & 16    & 12    \\
\hline 9    & 3    & 5    & 10    & 4    & 6    & 12    & 7    & 14    & 15    & 8    & 2    & 1    & 16    & 11    & 13    \\
\hline 10    & 9    & 4    & 6    & 8    & 1    & 7    & 16    & 15    & 14    & 12    & 5    & 3    & 2    & 13    & 11    \\
\hline 11    & 10    & 12    & 1    & 7    & 5    & 2    & 8    & 16    & 9    & 13    & 15    & 6    & 4    & 3    & 14    \\
\hline 12    & 11    & 10    & 15    & 2    & 8    & 6    & 3    & 5    & 16    & 14    & 13    & 9    & 7    & 1    & 4    \\
\hline 1    & 12    & 11    & 16    & 14    & 13    & 5    & 9    & 4    & 6    & 15    & 7    & 10    & 3    & 8    & 2    \\
\hline 3    & 2    & 13    & 11    & 12    & 14    & 9    & 6    & 8    & 1    & 7    & 10    & 16    & 15    & 4    & 5    \\
\hline 6    & 4    & 3    & 12    & 15    & 16    & 14    & 13    & 7    & 5    & 2    & 8    & 11    & 9    & 10    & 1    \\
\hline 13    & 7    & 1    & 4    & 16    & 12    & 15    & 14    & 2    & 8    & 6    & 3    & 5    & 11    & 9    & 10    \\
\hline 14    & 13    & 8    & 2    & 1    & 15    & 16    & 10    & 9    & 3    & 5    & 11    & 4    & 6    & 12    & 7    \\
\hline 15    & 14    & 16    & 5    & 3    & 2    & 13    & 11    & 10    & 12    & 4    & 6    & 8    & 1    & 7    & 9    \\
\hline 16    & 15    & 14    & 13    & 6    & 4    & 3    & 12    & 11    & 10    & 9    & 1    & 7    & 5    & 2    & 8    \\
\hline 5    & 16    & 15    & 14    & 13    & 7    & 1    & 4    & 12    & 11    & 10    & 9    & 2    & 8    & 6    & 3    \\
\hline
\end{tabular}
\label{example_thm6_ls2}
}
\caption{Example illustrating the ideas used in the proof of Theorem \ref{thm_completable_3}.}
\label{example_thm6}
\end{figure*}
Consider the singular fade state $s=\frac{\sin(\pi/16)}{\sin(2\pi/16)}e^{j \pi/16}$ for 16-PSK signal set. The constrained partial Latin Square for this case is shown in Fig. \ref{example_thm6_cpls}. Using the 8-coloring given in Lemma \ref{lemma9}, the partially filled Latin Square $L$ given in Fig. \ref{example_thm6_pfls} is obtained. 
Note that in $L,$ four cells are filled in every row and every column.
In the partially filled Latin Square $L,$ in each row, two more cells are filled in with the symbols 1 to 8, to obtain the partially filled Latin Square $L_1$ given in Fig. \ref{example_thm6_pfls_2}. The partially filled Latin Square $L'_1$ given in Fig. \ref{example_thm6_pfls_3} is obtained by interchanging the symbol and row indices of the filled cells in the partially filled Latin Square $L_1.$ Consider the partially filled Latin Rectangle $L'_R$ obtained by taking only the first 8 rows of $L'_1.$
Let $X_i, i \in \{1,2,\dotso,8\}$ denote the set of cells in which symbol $i$ can appear in $L'_R.$ Since two out of the 8 symbols $1,2,\dotso,8$ do not appear in the rows of $L_1,$ there are only two possibilities for the row indices of the cells in $X_i,$ which are denoted by $r_1(i)$ and $r_2(i).$ Let $X^j_i, j \in \{1,2\}$ denote those cells in $X_i$ which have row index $r_j(i).$ 
From Fig. \ref{example_thm6_pfls_3}, the sets $X^j_i, i \in \{1,2,\dotso,16\}, j \in \{1,2\},$ can be obtained as follows: 

{\scriptsize
\begin{tabular}{cc}
$X^1_1=\{(3,6),(3,10),(3,14)\},$ &$X^2_1=\{(7,4),(7,12),(7,16)\}$\\ $X^1_2=\{(4,7),(4,11),(4,15)\},$ & $X^2_2=\{(8,1),(8,5),(8,13)\}$\\ $X^1_3=\{(1,8),(1,12),(1,16)\},$& $X^2_3=\{(5,2),(5,6),(5,14)\},$\\ $X^1_4=\{(2,1),(2,9),(2,13)\},$ &$X^2_4=\{(6,3),(6,7),(6,15)\},$\\
$X^1_5=\{(3,2),(3,10),(3,14)\},$ &$X^2_5=\{(7,4),(7,8),(7,16)\},$\\
$X^1_6=\{(4,3),(4,11),(4,15)\},$ &$X^2_6=\{(8,1),(8,5),(8,9)\},$\\
$X^1_7=\{(1,4),(1,12),(1,16)\},$ &$X^2_7=\{(5,2),(5,6),(5,10)\},$\\
$X^1_8=\{(2,1),(2,5),(2,13)\},$ &$X^2_8=\{(6,3),(6,7),(6,11)\},$\\
$X^1_9=\{(3,2),(3,6),(3,14)\},$ &$X^2_9=\{(7,4),(7,8),(7,12)\},$\\
$X^1_{10}=\{(4,3),(4,7),(4,15)\},$ &$X^2_{10}=\{(8,5),(8,9),(8,13)\},$\\
$X^1_{11}=\{(1,4),(1,8),(1,16)\},$ &$X^2_{11}=\{(5,6),(5,10),(5,14)\},$\\
$X^1_{12}=\{(2,1),(2,5),(2,9)\},$ & $X^2_{12}=\{(6,7),(6,11),(6,15)\},$\\
$X^1_{13}=\{(3,2),(3,6),(3,10)\},$ &$X^2_{13}=\{(7,8),(7,12),(7,16)\},$\\
$X^1_{14}=\{(4,3),(4,7),(4,11)\},$ &$X^2_{14}=\{(8,1),(8,9),(8,13)\},$\\
$X^1_{15}=\{(1,4),(1,8),(1,12)\},$ &$X^2_{15}=\{(5,2),(5,10),(5,14)\},$\\
$X^1_{16}=\{(2,5),(2,9),(2,13)\},$ &$X^2_{16}=\{(6,3),(6,11),(6,15)\}.$
\end{tabular}
}

\noindent An SDR for $X=\{X^1_1,X^2_1,\dotso,X^1_{16},X^{2}_{16}\}$ is given by,

{\vspace{-.1 cm}
\scriptsize
\begin{align*}
&\left\lbrace (3,6),(7,4)(4,7),(8,1),(1,8),(5,2),(2,1)(6,3),(3,2),(7,8),(4,3),(8,5),\right.\\
&(1,4),(5,6),(2,5),(6,7),(3,14),(7,12),(4,15),(8,9),(1,16),(5,10),(2,9),\\
&\left.(6,11),(3,10),(7,16),(4,11),(8,13),(1,12),(5,14),(2,13),(6,15)\right\}.
\end{align*} 
\vspace{-.3 cm}
}
 
  \noindent The partially filled Latin Square in Fig. \ref{example_thm6_pfls_4} is obtained from $L'_R$ by filling the representative of $X^j_i$ with the symbol $i.$ 
From Theorem \ref{thm_Hall}, it follows that the partially filled Latin Square in Fig. \ref{example_thm6_pfls_4} is completable using $8$ symbols. Fig. \ref{example_thm6_ls1} shows one such completion. By interchanging the symbol and row indices in the Latin Square in Fig. \ref{example_thm6_ls1}, we get the Latin Square given in Fig. \ref{example_thm6_ls2} which removes the singular fade state $\frac{\sin(\pi/16)}{\sin(2\pi/16)}e^{j \pi/16}$ for 16-PSK signal set.
\end{example}
\section{Discussion}
It was shown that the Latin Squares which remove singular fade states for wireless two-way relaying can be obtained by coloring the vertices of the corresponding singularity removal graphs.  For $2^{\lambda}$-PSK signal set, $\lambda\geq 3,$ it was shown that the singularity removal graphs corresponding to all the singular fade states can be colored using $2^{\lambda}$ symbols. For $M$-QAM signal set, certain singular fade states for which the singularity removal graphs cannot be colored using $M$ colors were identified. The following are directions for future work:
\begin{itemize}
\item
It is possible to find a polynomial time algorithm to optimally color the singularity removal graphs for any arbitrary signal set?
\item
If the answer to the above question is no, for what class of signal sets is it possible to find a polynomial time algorithm?
\item 
For square QAM signal set, are there singular fade states other than the ones mentioned in Section IV, for which the singularity removal graphs cannot be colored using $M$ colors? 
\end{itemize}

\appendices
\section{Proof of Theorem \ref{thm4}:}
\label{appendix1}
The proof is provided for singular fade states of the form $\sin(\pi k/M)$ when $k$ is even and $\sin(\pi k/M)e^{j \pi/M}$ when $k$ is odd. The proof for the case when the singular fade state has absolute value $1/\sin(\pi k /M)$ follows, since a Latin Square which removes such a singular fade state can be obtained from a Latin Square which removes one of the singular fade states considered by taking transpose \cite{VNR}.

Let $S_i,i \in \{1,2,3,4\}$ denote the 4 sets mentioned in the order same as the one in the statement of Lemma \ref{lemma8} (Lemma \ref{lemma9}), for the case when $k$ is odd (even). Let $L$ denote the partially filled Latin Square obtained from the constrained partial Latin Square using the 4-coloring provided in Lemma \ref{lemma8} (Lemma \ref{lemma9}), when $k$ is odd (even) .
Let $i \in \{1,2,3,4\}$ be the symbol filled in the cells in $L$ which belong to the constraint $c_j,$ when $j$ belongs to the set $S_i.$
Note that $L$ contains 4 symbols, with each symbol appearing in $M/2$ cells and 2 cells filled in each row and column. 
 
The following approach is adopted to prove Theorem \ref{thm4}. First, a Latin Square $L_1$ and a Latin Rectangle $L'_R$ are obtained from $L.$ Next, some useful properties of $L_1$ and $L'_R$ are proved, before providing the proof of Theorem \ref{thm4}.

{\underline{\bf Obtaining $L_1$ and $L'_R$ from $L$:}}\\
\indent The Latin Square $L_1$ is obtained from $L$ by filling one more cell in every row from the symbol set $\{1,2,3,4\},$ without violating the exclusive law.\\ 
\mbox{\textit{\textbf{Case 1: $k$ is odd}}.}
For this case, it is claimed that the cells which belong to the set {\small$\left\{\left(i+1,\left(i-\frac{M}{4}-\frac{(k-1)}{2}\right)\text{mod}\:M+1\right), i \in \{0,1,\dotso,M-1\}\right\},$} are empty in $L.$ From \eqref{eqn3}, it follows that for $\left(i+1,\left(i-\frac{M}{4}-\frac{(k-1)}{2}\right)\text{mod}\:M+1\right)$ to be filled in $L,$ $i-\frac{M}{4}-\frac{(k-1)}{2}$ should be equal to either (i) $i-\frac{M}{4}-\frac{(k+1)}{2}$ or (ii) $i+\frac{M}{4}+\frac{(k-1)}{2}$ in modulo $M$ arithmetic. Satisfying the above requirement means either (i)$\frac{(k+1)}{2}=\frac{(k-1)}{2}$  (which is not possible), or (ii)$-k \:\text{mod}\: M=\frac{M}{2}+1$ (which is also not possible since $1\leq k\leq M/2-1$). 
Let $b_{i+1}$ denote the cell $\left(i+1,\left(i-\frac{M}{4}-\frac{(k-1)}{2}\right)\text{mod}\:M+1\right), i \in \{0,1,\dotso,M-1\}.$ We proceed to find which one of the four symbols 1, 2, 3 and 4 can be filled in $b_{i+1}.$ Consider the case when $i+1$ belongs to $S_1=\{1,3,\dotso,M/2-1\}.$ The singularity removal constraint $c_{i+1}$ has a cell with row index $i+1.$ Since the cells in $c_{i+1}$ are filled with 1 for $i+1 \in S_1,$ $b_{i+1}$ cannot be filled with 1. The only other singularity removal constraint which contains a cell with row index $i+1$ is $c_{i+k+1}.$ Since $i+k+1$ is even when $i+1 \in S_1,$ from the coloring provided in Lemma \ref{lemma8}, it follows that the cells which belong to $c_{i+k+1}$ are filled with either 2 or 4. Next, we need to look for filled cells which have column index $\left(i-\frac{M}{4}-\frac{(k-1)}{2}\right)\text{mod}\:M+1.$ The singularity removal constraints which contain cells with column index $\left(i-\frac{M}{4}-\frac{(k-1)}{2}\right)\text{mod}\:M+1$ are $c_{i+2}$ and $c_{i+M/2+2}.$ Since $i+2$ and $i+M/2+2$ are even for $i+1\in S_1,$ the cells in $c_{i+2}$ and $c_{i+M/2+2}$ are filled with either 2 or 4. Hence, for $i+1\in\{1,3,\dotso,M/2-1\},$ the cell $b_{i+1}$ can be filled with 3. 
Similarly, it can be shown that $b_{i+1}$ can be filled with the symbols 4, 1 and 2 when $i+1$ belongs to the sets $S_2,$ $S_3$ and $S_4$ respectively.  \\
\mbox{\textit{\textbf{Case 2: $k=a_1 2^m$ is even, $a_1$ is odd}}.}
It is claimed that the cells which belong to the set $\left(i+1,\left(i-\frac{M}{4}-\frac{k}{2}+2^{m}\right)\text{mod}\:M+1\right),$ $i \in \left\{0,1,\dotso,M-1\right\},$ are empty in $L.$ From \eqref{eqn1}, it follows that for $\left(i+1,\left(i-\frac{M}{4}-\frac{k}{2}+2^{m}\right)\text{mod}\:M+1\right)$ to be filled in $L,$ $i-\frac{M}{4}-\frac{k}{2}+2^m$ should be equal to either (i) $i-\frac{M}{4}-\frac{k}{2}$ or (ii) $i+\frac{M}{4}+\frac{k}{2}$ in modulo $M$ arithmetic, which are not possible. Let $b_{i+1}$ denote the cell $\left(i+1,\left(i-\frac{M}{4}-\frac{k}{2}+2^{m}\right)\text{mod}\:M+1\right).$ Similar to Case 1, it can be shown that, when $i+1$ belongs to $S_1,$ $S_2,$ $S_3$ and $S_4,$ the cell $b_{i+1}$ in $L$ can be filled with the symbols 3, 4, 1 and 2 respectively without violating the exclusive law.

 Note that in Case 1 and Case 2 above, the cell $b_{i+1}$ has row index $i+1$ and is filled in with the symbol which is the same as the one filled in the cells which belong to the singularity removal constraint $c_{(i+M/2)\text{mod}\:M+1}.$ Let $L_1$ denote the new partially filled Latin Square obtained after filling the cells $b_{i+1}, i+1 \in \{1,2,\dotso M\}$ in $L$ as mentioned in Case 1 and Case 2 above. Let $L'_1$ denote the partially filled Latin Square obtained by interchanging the symbol and row indices of the filled cells in $L_1.$ Since $L_1$ contains only 4 symbols, all the filled cells of $L'_1$ appear in the first 4 rows. 
Let $L'_R$ denote the partially filled Latin Rectangle formed by the first 4 rows of $L'_1.$

{\underline{\bf Some useful properties of $L_1$ and $L'_R$:}}\\
\indent We state some properties of $L_1$ and $L'_R$ which will be useful towards proving Theorem \ref{thm4}.

\textbf{Property 1:} Each one of the four symbols 1, 2, 3 and 4 appear in $3M/4$ columns in $L_1$ and hence there are $M/4$ empty cells in every row of $L'_R.$
  \begin{proof}
  Each one of the symbols  1, 2, 3 and 4 appear in $M/2$ cells in $L.$ This property follows from the fact that $L_1$ is obtained from $L$ by filling symbol $i\in \{1,2,3,4\}$ in $M/4$ empty cells of $L.$
  \end{proof}
  Let $\epsilon_1(i)$ denote the set of columns in which symbol $i$ is filled in $L'_R,$ which is the same as the set of columns which are filled in the $i$-th row of $L_1.$ Let $X_j$ denote the set of cells in $L'_R$ in which symbol $j\in \{1,2,\dotso,M\}$ can be filled without violating the exclusive law. Since every row of $L_1$ contains 3 filled cells, every symbol of $L'_R$ already appears in 3 rows. Hence the row indices of the cells which belong to $X_j$ are the same. Let $r(j)$ denote the row index of the cells which belong to $X_j.$ 

 \textbf{Property 2:} At least two of the three columns which are filled in the $j$-th row of $L_1$ contain $r(j).$
\begin{proof}
The proof is provided for the case when $k$ is odd. The idea used to prove this property for the case when $k$ is even is similar and is omitted.
Let $\kappa(j)$ denote the symbol filled in the cells which belong to the singularity removal constraint $c_{j}$ in $L_1.$ For $j = i+1\in \{1,2,\dotso,M\},$ the symbols which are filled in the $i+1$-th row of $L_1$ can be shown to be $\kappa(i+1),\kappa\left(\left(i+k\right)\text{mod}\:M+1\right)$ and $\kappa\left(\left(i+M/2)\text{mod}\:M+1\right)\right),$ which are respectively filled in the columns with indices $ (i-M/4-(k+1)/2)\text{mod}M+1,$ $(i+M/4+(k-1)/2)\text{mod}\:M+1$ and $(i-M/4-(k-1)/2)\text{mod}\:M+1.$ Consider the case when $i+1$ is odd. Since, $\left(i+M/2\right)\text{mod}\:M+1$ is also odd when $i+1$ is odd,  $\kappa(i+1)$ and $\kappa\left(\left(i+M/2\right)\text{mod}\:M+1\right)$ should be either 1 or 3. Hence $r(i+1)$ should be either 2 or 4. Consider the column with index $(i+M/4+(k-1)/2)\text{mod}\:M+1.$ In $L,$ the symbols filled in the column with index $(i+M/4+(k-1)/2)\text{mod}\:M+1$ can be shown to be $\kappa\left(\left(i+k\right)\text{mod}\:M+1\right)$ and $\kappa\left(\left(i+k-M/2\right)\text{mod}\:M+1\right).$ Since $\left(i+k\right)\text{mod}\:M+1$ and $\left(i+k-M/2\right)\text{mod}\:M+1$ are even when $i+1$ is odd, the two symbols filled in the column with index $(i+M/4+(k-1)/2)\text{mod}\:M+1$ in $L$ are 2 and 4. Hence, the column with index $(i+M/4+(k-1)/2)\text{mod}\:M+1$ in $L_1$ contains $r(i+1).$ The symbols which are filled in the $(i-M/4-(k-1)/2)\text{mod}\:M+1$-th column of $L_1$ can be shown to be $\kappa\left(\left(i+M/2\right)\text{mod}\:M+1\right),$ $\kappa\left(\left(i+1\right)\text{mod}\:M+1\right)$ and $\kappa\left(\left(i+M/2+1\right)\text{mod}\:M+1\right).$ Since $\left(i+1\right)\text{mod}\:M+1$ and $\left(i+M/2+1\right)\text{mod}\:M+1$ are even when $i+1$ is odd, $\kappa\left(\left(i+1\right)\text{mod}\:M+1\right)$ and $\kappa\left(\left(i+M/2+1\right)\text{mod}\:M+1\right)$ can only be 2 or 4. Hence, the column with index $(i-M/4-(k-1)/2)\text{mod}\:M+1$ in $L_1$ contains $r(i+1).$ Hence, at least two columns which are filled in the $i+1$-th row of $L_1$ contain $r(i+1).$ This completes the proof for the case when $j$ is odd. The proof for the case when $j$ is even is similar and is omitted.
\end{proof}    

For $i \in \{1,2,3,4\},$ let $S'_i$ denote the set $S'_i=\left\{j:r(j)=i,j\in\left\{1,2,\dotso,M\right\}\right\}.$ In other words, $S'_i$ denotes the set of rows in $L_1$ which do not contain symbol $i.$ Since every symbol from the set $\{1,2,3,4\}$ appears in $3M/4$ rows in $L_1,$ we have $\vert S'_i\vert=M/4, \forall i \in \{1,2,3,4\}.$\\
\textbf{Property 3:}
We have $\vert X_j\vert \geq \frac{M}{4}-1$ and $\displaystyle{\left\vert\bigcup_{j\in S'_i}X_j=\frac{M}{4}\right\vert},$ $\forall i \in \{1,2,3,4\}.$
\begin{proof}
Since every symbol appears in $3M/4$ columns in $L_1,$ there are $M/4$ empty cells in the $r(j)$-th row of $L'_R.$ Hence $\vert X_j \vert$ can be at most $M/4.$ It is argued out that $\vert X_j \vert$ should be at least $M/4-1.$ Let $\epsilon_2(j)$ denote the columns which are empty in the $r(j)$-th row in $L'_R$ which are the same as the columns in which symbol $r(j)$ is not filled in $L_1.$ To show that $\vert X_j \vert$ should be at least $M/4-1,$ it suffices to show that $\vert \epsilon_2(j) \cap \epsilon_1(j)\vert \leq 1,$ which boils down to showing that at least two of the three columns which belong to $\epsilon_1(j)$ contain the symbol $r(j)$ in $L_1,$ which follows from Property 2. 

Since $\vert X_j\vert \geq \frac{M}{4}-1,$ $\displaystyle{\left\vert\bigcup_{j\in S'_i}X_j\right\vert<\frac{M}{4}}$ is possible only when there exists an empty cell in $i$-th row of $L'_R$ in which none of the $M/4$ symbols which belong to the set $S'_i$ can be filled, i.e., the column in $L'_R$ in which that cell lies contains all the elements of the set $S'_i.$ For $M\geq 16,$ such an empty cell cannot exist since, $M/4\geq 4,$ while the number of filled cells in a column is only three. For $M=8,$ the validity of this result has been verified by brute force. 
\end{proof}  

Now, we proceed to give the proof of Theorem \ref{thm4}.

\textbf{\underline{Proof of Theorem \ref{thm4}}:}\\
\indent If $L'_R$ is completable using $M$ symbols, then from Theorem \ref{thm_Hall} it would mean that $L'_1$ is completable using $M$ symbols, which would in turn mean that $L_1$ is completable using $M$ symbols. Hence, to complete the proof, it suffices to show that $L'_R$ is completable using $M$ symbols. 



To show that $L'_R$ is completable, it suffices to show that an SDR exists for $X=\{X_1,X_2,\dotso X_M\}.$ A completion of $L'_R$ using $M$ symbols can be obtained by filling the representative of $X_i$ with the symbol $i$ in $L'_R.$ 

For $i \in \{1,2,3,4\},$ let $s_i \subseteq S'_i.$ To show that $X$ has an SDR, from Theorem \ref{thm_marriage}, it needs to be shown that $\displaystyle{\left\vert\bigcup_{i \in \{1,2,3,4\}}\bigcup_{j \in s_i}X_j\right\vert \geq \sum_{i=1}^{4}\vert s_i\vert.}$ Note that for $i\in S'_{l_1}$ and $j \in S'_{l_2},$ where $l_1 \neq l_2,$ $r(i)\neq r(j).$ Hence, we have, $\displaystyle{\left\vert\bigcup_{i \in \{1,2,3,4\}}\bigcup_{j \in s_i}X_j\right\vert=\sum_{i=1}^4 \left\vert\bigcup_{j \in s_i}X_j\right\vert}.$ 
From Property 3 and from the fact that $\vert S'_i \vert =M/4,$ it follows that  $\displaystyle{\left\vert\bigcup_{j \in s_i}X_j\right\vert\geq \vert s_i\vert}.$  Hence, we have  $\displaystyle{\left\vert\bigcup_{i \in \{1,2,3,4\}}\bigcup_{j \in s_i}X_j\right\vert \geq \sum_{i=1}^{4}\vert s_i\vert.}$ This completes the proof of Theorem \ref{thm4}.
\endproof
\section{Proof of Theorem \ref{thm_completable_3}:}
\label{appendix2}
The approach adopted to prove Theorem \ref{thm_completable_3} is very similar to the one used in the proof of Theorem \ref{thm4} in Appendix \ref{appendix1}.
Let $S_i,i \in \{1,2,\dotso,8\}$ denote the 8 sets mentioned in the order same as the one in the statement of Lemma \ref{lemma6} (Lemma \ref{lemma7}), for the case when both $k$ and $l$ are even (only one among $k$ and $l$ is even), where $k,l \neq M/2.$ Let $L$ denote the corresponding partially filled Latin Square obtained from the constrained partial Latin Square using the 8-coloring provided in Lemma \ref{lemma6} (Lemma \ref{lemma7}).
Let $i \in \{1,2,\dotso,8\}$ be the symbol filled in the cells in $L$ which belong to the constraint $c_j,$ when $j$ belongs to the set $S_i.$
Note that $L$ contains 8 symbols, with each symbol appearing in $M/2$ cells and 4 cells filled in each row and column. 

In the first step, two empty cells are filled in every row in $L$ to get a partially filled Latin Square $L_1$ with six filled cells in every row and every column using the procedure described below. An $8\times M$ Latin Rectangle $L'_R$ is obtained from $L_1$ by interchanging the symbol and row indices of the filled cells in $L_1.$
 
{\underline{\bf Obtaining $L_1$ and $L'_R$ from $L$:}}\\
\mbox{\textit{\textbf{Case 1: only one among $k$ and $l$ is even}}.}
Let $b^1_i$ denote the cell $\left(i+1,\left(i-\frac{k+1-l}{2}\right)\text{mod}\:M+1\right),$ where $i\in\{0,1,\dotso,M-1\}.$ It is claimed that the cell $b^1_i$ is empty in $L,$ for all $i \in \{0,1,\dotso M-1\}.$ Otherwise, from \eqref{eqn3} and \eqref{eqn4}, $i-\frac{k+1-l}{2}$ should equal (i) $i-\frac{M}{2}-\frac{k+1-l}{2}$  or (ii) $i+\frac{M}{2}+\frac{k-l-1}{2}$  or (iii) $i-\frac{k+l+1}{2}$ or (iv) $i+\frac{k+l-1}{2}$. It can be verified that none of the four conditions mentioned above are possible.
The cell $b^1_i$ in $L$ can be filled in with a symbol from the set $\{1,2,3,4\}$ without violating the exclusive law. The reason for this is explained below. Let $\zeta(u,v)$ denote the union of the set of symbols from the set $\{1,2,3,4\}$ which are filled in the $u$-th row and $v$-th column. If $(u,v)$ is a filled cell in $L,$ the cardinality of $\zeta(u,v)$ is three, since in every row (column), two cells are filled with symbols from the set $\{1,2,3,4\}.$   
Since, the vertices which differ by $M/2$ are colored using the same color in the coloring provided in Lemma \ref{lemma7}, the set $\zeta(b^1_i)$ is the same as the set $\zeta\left(i+1,\left(i+M/2-\frac{k+1-l}{2}\right)\text{mod}\:M+1\right).$ Since $\left(i+1,\left(i+M/2-\frac{k+1-l}{2}\right)\text{mod}\:M+1\right)$ is a filled cell in $L,$ $\zeta(b^1_i)$ is of cardinality three. The symbol from the set $\{1,2,3,4\}$ which does not belong to $\zeta(b^1_i)$ can be filled in $b^1_i.$
Let $b^2_i$ denote the cell $(i+1,\left(i+\frac{M}{2}-\frac{k+1+l}{2}\right)\text{mod}\:M+1),$ where $i\in\{0,1,\dotso,M-1\}.$ Using a similar argument as above, it can be shown that the cell $b^2_i$ can be filled in with a symbol from the set $\{5,6,7,8\}$ without violating the exclusive law.\\
\mbox{\textit{\textbf{Case 2: both $k$ and $l$ are even}}.}
Let $b^1_i$ denote the cell $\left(i+1,\left(i-\frac{k-l}{2}\right)\text{mod}\:M+1\right),$ where $i\in\{0,1,\dotso,M-1\}.$ Let $b^2_i$ denote the cell $(i+1,\left(i+\frac{M}{2}-\frac{k+l}{2}\right)\text{mod}\:M+1),$ where $i\in\{0,1,\dotso,M-1\}.$ Similar to Case 1, it can be shown that the cell $b^1_i$ ($b^2_i$) can be filled in with a symbol from the set $\{1,2,3,4\}$ ($\{5,6,7,8\}$) without violating the exclusive law.

Let $L_1$ be the partially filled Latin Square obtained from $L$ by filling the cells $b^1_i$ and $b_2^i$ as described above. Let $L'_1$ denote the Latin Square obtained by interchanging the symbol and row indices of the filled cells of $L_1.$ Since $L_1$ is filled with only 8 symbols, only the first 8 rows of $L'_1$ contain filled cells. Let $L'_R$ denote the Latin Rectangle obtained by taking the first 8 rows of $L'_R.$

{\underline{\bf Some useful properties of $L_1$ and $L'_R$:}}\\
\indent Let $X_i$ denote the set of cells in $L'_R$ in which symbol $i$ can be filled without violating the exclusive law. Since 6 cells are filled in the $i^{th}$ row of $L_1,$ there are only two possibilities for the row indices for the cells which belong to $X_i,$ denoted by $r_1(i)$ and $r_2(i).$ Let $X^1_i$ and $X^2_i$ denote the set of cells in $X(i)$ which have its row indices to be $r_1(i)$ and $r_2(i)$ respectively. There are $2M$ empty cells in $L'_R$ in total. Every symbol $i$ needs to be filled in 2 of those $2M$ cells, with one of the 2 cells belonging to $X^1_i$ and the other one belonging to $X^2_i.$

\textbf{Property 1:}
The values of $r_1(i)$ and $r_2(i)$ differ by 4.
\begin{proof}
Let $\kappa(j)$ denote the symbols filled in the cells which belong to the singularity removal constraint $c_j.$
 The symbols which are not filled in the $i+1$-th row of $L_1$ can be shown to be  $\kappa(c_{(i+l)\text{mod}\:M})$ and $\kappa(c_{(i+l)\text{mod}\:M+M}).$ Note that according to the coloring provided in Lemma \ref{lemma6} and Lemma \ref{lemma7}, the cells which belong to two constraints $c_{u_1}$ and $c_{u_2},$ where $u_1$ and $u_2$ differ by $M,$ are filled in with symbols which differ by 4. Hence, $r_1(i)$ and $r_2(i)$ differ by 4.
\end{proof}
\textbf{Property 2:}
In $L_1,$ if a symbol $j\leq4$ ($j>4$) does not appear in a column, then the symbol $j+4$ ($j-4$) appears in that column.
\begin{proof}
The proof is provided for the case when only one among $k$ and $l$ is even. The proof for the case when both are even is similar and is omitted.
Let $\kappa(j)$ denote the symbols filled in the cells which belong to the singularity removal constraint $c_j.$
The symbols from $\{1,2,3,4\}$ which are filled in the $\left(i-\frac{M}{2}-\frac{k+1-l}{2}\right)\text{mod}\:M+1$-th column are given by,

{\footnotesize
\begin{align*}
&\kappa(i+1),\kappa\left(\left(i+l\right)\text{mod}\:M+1\right)\:\text{and} \\
&\{1,2,3,4\}\setminus\left\{\kappa(i+1),\kappa\left(\left(i+l\right)\text{mod}\:M+1\right),\kappa\left(\left(i+k\right)\text{mod}\:M+1\right)\right\}.
\end{align*}}

\noindent The symbols from $\{5,6,7,8\}$ which are filled in the $\left(i-\frac{M}{2}-\frac{k+1-l}{2}\right)\text{mod}\:M+1$-th column are given by,

{\footnotesize
\begin{align*}
&\kappa(i+1+M),\kappa\left(\left(i+l\right)\text{mod}\:M+1+M\right)\:\text{and} \\
&\{5,6,7,8\}\setminus\left\{\kappa(i+1+M),\kappa\left(\left(i+l\right)\text{mod}\:M+1+M\right),\right.\\
&\hspace{4cm}\left.\kappa\left(\left(i+k-l\right)\text{mod}\:M+1+M\right)\right\}.
\end{align*}}

\noindent The result follows from the fact that $\kappa\left(\left(i+k\right)\text{mod}\:M+1\right)$ and $\kappa\left(\left(i+k-l\right)\text{mod}\:M+1+M\right)$ do not differ by 4.
\end{proof}
\textbf{Property 3:} For every $i,$  the set of column indices of the cells which belong to $X^1_i$ has no intersection with the set of column indices of the cells which belong to $X^2_i.$
\begin{proof}
Towards proving this, it suffices to show that the set of column indices corresponding to the empty cells in row $r_1(i)$ has no intersection with the set of column indices corresponding to the empty cells in row $r_2(i)$ in $L'_R.$ From Property 1 and Property 2, it follows that the set of column indices corresponding to the empty cells in row $r_1(i)$ has no intersection with the set of column indices corresponding to the empty cells in row $r_2(i).$
\end{proof}

From Property 1, it follows that only one among $r_1(i)$ and $r_2(i)$ is less than or equal to 4. Without loss of generality assume that $r_1(i)\leq 4$ and $r_2(i)> 4.$

Let $\epsilon_1(i)$ denote the set of columns in which symbol $i$ is filled in $L'_R,$ which is the same as the set of columns which are filled in the $i$-th row of $L_1.$ 

 \textbf{Property 4:} There is only one column which is filled in the $j$-th row of $L_1,$ which 
 does not contain $r_1(j).$
\begin{proof}
Let $\kappa(j)$ denote the symbol filled in the cells which belong to the singularity removal constraint $c_{j}$ 
in $L_1.$ Let $z^{1}_q(i)$ ( $z^{2}_r(i)$) $q \in \{1,2,\dotso 6\}$ denote the six sets of symbols from the set $\{1,2,3,4\}$ ($\{5,6,7,8\})$) which are filled in the columns which belong to $\epsilon_1(i)$ in $L_1.$ Note that $\vert z^{e}_q(i)\vert =3,$ $\forall e\in \{1,2\}, q \in \{1,2,\dotso,6\}.$

Let $k=2^{m_1}a_1$ and $l=2^{m_2}a_2,$ $a_1$ and $a_2$ being odd with $m_1<m_2.$ 
From the colorings provided in \ref{lemma6} and Lemma \ref{lemma7}, it can be shown that $r_1(i+1)=\kappa\left(\left(i+2^{m_2}\right)\text{mod}\:M+1\right),$ for $i+1 \in \{1,2,\dotso,M\}.$ 
Also, the six sets $z^{1}_q(i),$ $q \in \{1,2,\dotso 6\}$ can be shown to be 

{\scriptsize
\begin{align*}
&z^{1}_1(i+1)=z^{1}_{3}(i+1)=\left\{\kappa(i+2^{m_2}),\kappa\left(\left(i+2^{m_2}\right)\text{mod}\:M+1\right),\right.\\ 
&\hspace{3 cm}\left.\kappa\left(\left(i+2^{m_2}+2^{m_1}\right)\text{mod}\:M+1\right)\right\},\\
&z^{1}_2(i+1)=\left\{\kappa(i+2^{m_1}),\kappa\left(\left(i+2^{m_2}\right)\text{mod}\:M+1\right),\right.\\
&\hspace{3 cm} \left.\kappa\left(\left(i+2^{m_2}+2^{m_1}\right)\text{mod}\:M+1\right)\right\},\\
&z^{1}_4(i+1)=z^{1}_6(i+1)=\left\{\kappa(i+1),\kappa\left(\left(i+2^{m_1}\right)\text{mod}\:M+1\right)\right.,\\
&\hspace{3 cm}\left. \kappa\left(\left(i+2^{m_2}\right)\text{mod}\:M+1\right)\right\},\\
&z^{1}_5(i+1)=\left\{\kappa(i+1),\kappa\left(\left(i+2^{m_1}\right)\text{mod}\:M+1\right),\right.\\
&\hspace{3 cm} \left.\kappa\left(\left(i+2^{m_2}+2^{m_1}\right)\text{mod}\:M+1\right)\right\}.
\end{align*}}

Since five of the six sets given above contain $\kappa\left(\left(i+2^{m_2}\right)\text{mod}\:M+1\right),$ there is only one column which is filled in the $j$-th row of $L_1,$ which 
 does not contain $r_1(j).$
\end{proof} 

For $i\in\{1,2,3,4\},$ let $S'_i$ denote the set $\{j:r_1(j)=i, j \in \{1,2,\dotso M\}\}.$ Similarly, for $i\in\{5,6,7,8\},$ let $S'_i$ denote the set $\{j:r_2(j)=i, j \in \{1,2,\dotso M\}\}.$ Since every symbol appears in $3M/4$ rows in $L_1,$ we have $\vert S'_i\vert=M/4, \forall i \in \{1,2,\dotso,8\}.$\\

\textbf{Property 5:}
We have  \mbox{$\vert X^e_j\vert \geq \frac{M}{4}-1$} and \mbox{$\displaystyle{\left\vert\bigcup_{j \in S'_i}X_j^e\right\vert\geq \frac{M}{4}}$}, for all $e\in \{1,2\}$ and $i\in\{1,2,\dotso,8\}.$
\begin{proof}
Let $\epsilon_2(j)$ denote the set of columns which are empty in the $r_1(j)$-th row of $L'_R$ which is the same as the set of columns in which symbol $r_1(j)$ is not filled in $L_1.$ Note that $\vert \epsilon_2(i) \vert =M/4.$ Since $X^1_j$ is the set of cells in the $r_1(j)$-th row of $L'_R$ in which symbol $j$ can be filled, it equals $\epsilon_2(j)\setminus \epsilon_1(j).$ Hence, to show that $\vert X^1_j\vert \geq M/4-1,$ it suffices to show that $\vert \epsilon_2(j) \cap \epsilon_1(j)\vert \leq 1.$ Note that $\epsilon_2(j) \cap \epsilon_1(j)$ is the set of columns which are filled in the $j$-th row of $L_1$ which do not contain the symbol $r_1(j).$ Hence, from Property 4, we have, $\vert \epsilon_2(j) \cap \epsilon_1(j)\vert =1.$ This completes the proof that $\vert X^1_j\vert \geq \frac{M}{4}-1.$ By a similar reasoning, we have, $\vert X^2_j\vert \geq \frac{M}{4}-1.$

Now, we proceed to show that  \mbox{$\displaystyle{\left\vert\bigcup_{j \in S'_i}X_j^i\right\vert\geq \frac{M}{4}}.$} On the contrary, assume \mbox{$\displaystyle{\left\vert\bigcup_{j \in S'_i}X_j^1\right\vert< \frac{M}{4}}.$} Since, \mbox{$\vert X^1_j\vert \geq \frac{M}{4}-1$}, in that case, all the sets $X^1_j,j \in S'_i,$ have to be equal with cardinality $M/4-1.$ This means that there should exists an empty cell in the $r_1(j)$-th row of $L'_R$ in which all the $M/4$ symbols which belong to $S'_i$ can not appear. For $M\geq 32,$ such an empty cell cannot exist since, $M/4\geq 8,$ while the number of filled cells in a column is only six. For $M=8$ and $M=16,$ the validity of this result has been verified by brute force. 
\end{proof}
\textbf{\underline{Proof of Theorem \ref{thm_completable_3}:}}\\
From Property 3, it follows that the set of column indices of the cells which belong to $X^1_i$ has no intersection with the set of column indices of the cells which belong to $X^2_i.$ Hence, to show the completability of $L'_R$ using $M$ symbols, it suffices to show that an SDR exists for the set $X=\{X^1_1,X^2_1, X^1_2,X^2_2,\dotso, X^1_M,X^2_M\}.$ 

Let $s_i$ denote a subset of $S'_i,$ where $i \in \{1,2,\dotso,8\}.$ To show the existence of SDR for $X$ it needs to be shown that $\displaystyle{\left\vert\left(\bigcup_{i \in \{1,2,3,4\}}\bigcup_{j \in s_i}X^1_{j}\right)\cup\left(\bigcup_{i \in \{5,6,7,8\}}\bigcup_{j \in s_i}X^2_{j}\right)\right\vert \geq \sum_{i=1}^{8}\vert s_i\vert.}$

For $j_1 \in s_{i_1},j_2 \in s_{i_2},$ $i_1\neq i_2$ and $i_1,i_2\leq 4$ ($i_1,i_2 >4$), the row indices of the cells in $X^1_{j_1}$ and $X^1_{j_2}$ ($X^2_{j_1}$ and $X^2_{j_2}$) are different.
Similarly, for $j_1 \in s_{i_1},j_2 \in s_{i_2},$ $i_1\neq i_2,$ $i_1\leq4$ and $i_2> 4$ ($i_1>4 $ and $i_2\leq4$), the row indices of the cells in $X^1_{j_1}$ and $X^2_{j_2}$ ($X^2_{j_1}$ and $X^1_{j_2}$) are different.
 Hence $X^e_{j_1}$ and $X^e_{j_2}$ have no element in common, for $e\in \{1,2\},$ $j_1\in s_{i_1},j_2\in s_{i_2}$ and $i_1 \neq i_2.$
From the above fact, we have,

{\scriptsize
\begin{align*}
&\left\vert\left(\bigcup_{i \in \{1,2,3,4\}}\bigcup_{j \in s_i}X^1_{j}\right)\cup\left(\bigcup_{i \in \{5,6,7,8\}}\bigcup_{j \in s_i}X^2_{j}\right)\right\vert\\
&\hspace{4cm}=\sum_{i =1}^4\left\vert\bigcup_{j \in s_i} X^1_{j}\right\vert +\sum_{i =5}^8\left\vert\bigcup_{j \in s_i} X^2_{j}\right\vert.
\end{align*}}
From the facts that \mbox{$\vert S_i \vert =\frac{M}{4},$} \mbox{$\vert X^e_j\vert \geq \frac{M}{4}-1$} and \mbox{$\displaystyle{\left\vert\bigcup_{j \in S_i}X_j^e\right\vert\geq \frac{M}{4}}$} (Property 5), it follows that \mbox{$\displaystyle{\left\vert\bigcup_{j \in s_i} X^e_{j}\right\vert\geq \vert s_i \vert.}$} Hence, we have, $${\left\vert\left(\bigcup_{i \in \{1,2,3,4\}}\bigcup_{j \in s_i}X^1_{j}\right)\cup\left(\bigcup_{i \in \{5,6,7,8\}}\bigcup_{j \in s_i}X^2_{j}\right)\right\vert \geq \sum_{i=1}^{8}\vert s_i\vert.}$$

\noindent This completes the proof of Theorem \ref{thm_completable_3}.
\endproof

\end{document}